\documentclass[twocolumn, appendixfloats, revtex4, numberedappendix, twocolappendix, iop, letterpaper]{openjournal}
\usepackage[pass]{geometry}

\usepackage{natbib}

\usepackage[breaklinks,colorlinks,linkcolor=blue,citecolor=blue,urlcolor=blue]{hyperref}

\usepackage[T1]{fontenc}

\usepackage{orcidlink}

\usepackage{xspace}

\usepackage{graphicx}

\usepackage{paralist}

\newcommand{\HI}{\ifmmode \mathrm{H}\,\textsc{i} \else H~{\sc i}\fi\xspace}
\newcommand{\nhi}{\ifmmode \mathrm{N(H}\,\textsc{i}) \else N(H~{\sc i})\fi\xspace}
\newcommand{\mgii}{\ifmmode \mathrm{Mg}\,\textsc{ii} \else Mg~{\sc ii}\fi\xspace}
\newcommand{\civ}{\ifmmode \mathrm{C}\,\textsc{iv} \else C~{\sc iv}\fi\xspace}
\newcommand{\ciii}{\ifmmode \mathrm{C}\,\textsc{iii}] \else C~{\sc iii}]\fi\xspace}
\newcommand{\mgal}{\ifmmode M_* \else $M_*$\fi\xspace}
\newcommand{\modot}{\ifmmode M_\odot \else $M_\odot$\fi\xspace}

\shorttitle{Cool Gas in the CGM of Post Starburst Galaxies}
\shortauthors{Z. Harvey et al.}

\begin{document}

\title{The Evolution of the Circumgalactic Medium after a starburst in DESI Post-Starburst Galaxies\vspace{-0.8cm}\vspace{-0.7cm}}

\author{Zoe Harvey\,\orcidlink{0009-0009-9830-550X}$^{1}$,
  Sahyadri Krishna\,\orcidlink{0009-0000-5833-4116}$^{1}$,
  Rita Tojeiro\,\orcidlink{0000-0001-5191-2286}$^{1}$,
  Vivienne Wild\,\orcidlink{0000-0002-8956-7024}$^{1}$,
  Jessica Nicole Aguilar$^{2}$,
  Steven Ahlen\,\orcidlink{0000-0001-6098-7247}$^{3}$,
  Eric Armengaud\,\orcidlink{0000-0001-7600-5148}$^{4}$,
  Florian Beutler\,\orcidlink{0000-0003-0467-5438}$^{5}$,
  Davide Bianchi\,\orcidlink{0000-0001-9712-0006}$^{6,7}$,
  David Brooks$^{8}$,
  Rebecca Canning$^{9}$,
  Aurelio Carnero Rosell\,\orcidlink{0000-0003-3044-5150}$^{10,11}$,
  Todd Claybaugh$^{2}$,
  Andrei Cuceu\,\orcidlink{0000-0002-2169-0595}$^{2}$,
  Axel de la Macorra\,\orcidlink{0000-0002-1769-1640}$^{12}$,
  Peter Doel$^{8}$,
  Jaime E. Forero-Romero\,\orcidlink{0000-0002-2890-3725}$^{13,14}$,
  Enrique Gazta\~{n}aga\,\orcidlink{0000-0001-9632-0815}$^{9,15,16}$,
  Satya Gontcho A Gontcho\,\orcidlink{0000-0003-3142-233X}$^{17}$,
  Gaston Gutierrez$^{18}$,
  ChangHoon Hahn\,\orcidlink{0000-0003-1197-0902}$^{19}$,
  Hiram K. Herrera-Alcantar\,\orcidlink{0000-0002-9136-9609}$^{4,20,21,22,23}$,
  Stephanie Juneau\,\orcidlink{0000-0002-0000-2394}$^{24}$,
  Robert Kehoe$^{25}$,
  Martin Landriau\,\orcidlink{0000-0003-1838-8528}$^{2}$,
  Laurent Le Guillou\,\orcidlink{0000-0001-7178-8868}$^{26}$,
  Marc Manera\,\orcidlink{0000-0003-4962-8934}$^{27,28}$,
  Aaron Meisner\,\orcidlink{0000-0002-1125-7384}$^{24}$,
  Ramon Miquel$^{28,29}$,
  John Moustakas\,\orcidlink{0000-0002-2733-4559}$^{30}$,
  Adam Myers$^{31}$,
  Seshadri Nadathur\,\orcidlink{0000-0001-9070-3102}$^{9}$,
  Jeffrey A. Newman\,\orcidlink{0000-0001-8684-2222}$^{32}$,
  Enrique Paillas\,\orcidlink{0000-0002-4637-2868}$^{33}$,
  Will Percival\,\orcidlink{0000-0002-0644-5727}$^{34,35,36}$,
  Ignasi P\'{e}rez-R\`{a}fols\,\orcidlink{0000-0001-6979-0125}$^{37}$,
  Francisco Prada\,\orcidlink{0000-0001-7145-8674}$^{38}$,
  Corentin Ravoux\,\orcidlink{0000-0002-3500-6635}$^{39}$,
  Graziano Rossi$^{40}$,
  Rossana Ruggeri\,\orcidlink{0000-0002-0394-0896}$^{41}$,
  Eusebio Sanchez\,\orcidlink{0000-0002-9646-8198}$^{42}$,
  Christoph Saulder\,\orcidlink{0000-0002-0408-5633}$^{43}$,
  David Schlegel$^{2}$,
  Joseph Harry Silber\,\orcidlink{0000-0002-3461-0320}$^{2}$,
  Ma\l{}gorzata Siudek\,\orcidlink{0000-0002-2949-2155}$^{11,16}$,
  Gregory Tarl\'{e}\,\orcidlink{0000-0003-1704-0781}$^{44}$,
  Benjamin Alan Weaver$^{24}$,
  Hu Zou\,\orcidlink{0000-0002-6684-3997}$^{45}$\\
}

\affiliation{
  $^{1}$School of Physics and Astronomy, University of St Andrews, St Andrews, Fife KY16 8AZ\\
  $^{2}$Lawrence Berkeley National Laboratory, 1 Cyclotron Road, Berkeley, CA 94720, USA\\
  $^{3}$Department of Physics, Boston University, 590 Commonwealth Avenue, Boston, MA 02215 USA\\
  $^{4}$IRFU, CEA, Universit\'{e} Paris-Saclay, F-91191 Gif-sur-Yvette, France\\
  $^{5}$Institute for Astronomy, University of Edinburgh, Royal Observatory, Blackford Hill, Edinburgh EH9 3HJ, UK\\
  $^{6}$Dipartimento di Fisica ``Aldo Pontremoli'', Universit\`a degli Studi di Milano, Via Celoria 16, I-20133 Milano, Italy\\
  $^{7}$INAF-Osservatorio Astronomico di Brera, Via Brera 28, 20122 Milano, Italy\\
  $^{8}$Department of Physics \& Astronomy, University College London, Gower Street, London, WC1E 6BT, UK\\
  $^{9}$Institute of Cosmology and Gravitation, University of Portsmouth, Dennis Sciama Building, Portsmouth, PO1 3FX, UK\\
  $^{10}$Departamento de Astrof\'{\i}sica, Universidad de La Laguna (ULL), E-38206, La Laguna, Tenerife, Spain\\
  $^{11}$Instituto de Astrof\'{\i}sica de Canarias, C/ V\'{\i}a L\'{a}ctea, s/n, E-38205 La Laguna, Tenerife, Spain\\
  $^{12}$Instituto de F\'{\i}sica, Universidad Nacional Aut\'{o}noma de M\'{e}xico,  Circuito de la Investigaci\'{o}n Cient\'{\i}fica, Ciudad Universitaria, Cd. de M\'{e}xico  C.~P.~04510,  M\'{e}xico\\
  $^{13}$Departamento de F\'isica, Universidad de los Andes, Cra. 1 No. 18A-10, Edificio Ip, CP 111711, Bogot\'a, Colombia\\
  $^{14}$Observatorio Astron\'omico, Universidad de los Andes, Cra. 1 No. 18A-10, Edificio H, CP 111711 Bogot\'a, Colombia\\
  $^{15}$Institut d'Estudis Espacials de Catalunya (IEEC), c/ Esteve Terradas 1, Edifici RDIT, Campus PMT-UPC, 08860 Castelldefels, Spain\\
  $^{16}$Institute of Space Sciences, ICE-CSIC, Campus UAB, Carrer de Can Magrans s/n, 08913 Bellaterra, Barcelona, Spain\\
  $^{17}$University of Virginia, Department of Astronomy, Charlottesville, VA 22904, USA\\
  $^{18}$Fermi National Accelerator Laboratory, PO Box 500, Batavia, IL 60510, USA\\
  $^{19}$Department of Astronomy, University of Texas at Austin, 2515 Speedway, TX 78712, USA\\
  $^{20}$Institut d'Astrophysique de Paris. 98 bis boulevard Arago. 75014 Paris, France\\
  $^{21}$Center for Cosmology and AstroParticle Physics, The Ohio State University, 191 West Woodruff Avenue, Columbus, OH 43210, USA\\
  $^{22}$Department of Astronomy, The Ohio State University, 4055 McPherson Laboratory, 140 W 18th Avenue, Columbus, OH 43210, USA\\
  $^{23}$The Ohio State University, Columbus, 43210 OH, USA\\
  $^{24}$NSF NOIRLab, 950 N. Cherry Ave., Tucson, AZ 85719, USA\\
  $^{25}$Department of Physics, Southern Methodist University, 3215 Daniel Avenue, Dallas, TX 75275, USA\\
  $^{26}$Sorbonne Universit\'{e}, CNRS/IN2P3, Laboratoire de Physique Nucl\'{e}aire et de Hautes Energies (LPNHE), FR-75005 Paris, France\\
  $^{27}$Departament de F\'{i}sica, Serra H\'{u}nter, Universitat Aut\`{o}noma de Barcelona, 08193 Bellaterra (Barcelona), Spain\\
  $^{28}$Institut de F\'{i}sica d'Altes Energies (IFAE), The Barcelona Institute of Science and Technology, Edifici Cn, Campus UAB, 08193, Bellaterra (Barcelona), Spain\\
  $^{29}$Instituci\'{o} Catalana de Recerca i Estudis Avan\c{c}ats, Passeig de Llu\'{\i}s Companys, 23, 08010 Barcelona, Spain\\
  $^{30}$Department of Physics and Astronomy, Siena University, 515 Loudon Road, Loudonville, NY 12211, USA\\
  $^{31}$Department of Physics \& Astronomy, University  of Wyoming, 1000 E. University, Dept.~3905, Laramie, WY 82071, USA\\
  $^{32}$Department of Physics \& Astronomy and Pittsburgh Particle Physics, Astrophysics, and Cosmology Center (PITT PACC), University of Pittsburgh, 3941 O'Hara Street, Pittsburgh, PA 15260, USA\\
  $^{33}$Instituto de Estudios Astrof\'isicos, Facultad de Ingenier\'ia y Ciencias, Universidad Diego Portales, Av. Ej\'ercito Libertador 441, Santiago, Chile\\
  $^{34}$Department of Physics and Astronomy, University of Waterloo, 200 University Ave W, Waterloo, ON N2L 3G1, Canada\\
  $^{35}$Perimeter Institute for Theoretical Physics, 31 Caroline St. North, Waterloo, ON N2L 2Y5, Canada\\
  $^{36}$Waterloo Centre for Astrophysics, University of Waterloo, 200 University Ave W, Waterloo, ON N2L 3G1, Canada\\
  $^{37}$Departament de F\'isica, EEBE, Universitat Polit\`ecnica de Catalunya, c/Eduard Maristany 10, 08930 Barcelona, Spain\\
  $^{38}$Instituto de Astrof\'{i}sica de Andaluc\'{i}a (CSIC), Glorieta de la Astronom\'{i}a, s/n, E-18008 Granada, Spain\\
  $^{39}$Universit\'{e} Clermont-Auvergne, CNRS, LPCA, 63000 Clermont-Ferrand, France\\
  $^{40}$Department of Physics and Astronomy, Sejong University, 209 Neungdong-ro, Gwangjin-gu, Seoul 05006, Republic of Korea\\
  $^{41}$Queensland University of Technology,  School of Chemistry \& Physics, George St, Brisbane 4001, Australia\\
  $^{42}$CIEMAT, Avenida Complutense 40, E-28040 Madrid, Spain\\
  $^{43}$Max Planck Institute for Extraterrestrial Physics, Gie\ss enbachstra\ss e 1, 85748 Garching, Germany\\
  $^{44}$University of Michigan, 500 S. State Street, Ann Arbor, MI 48109, USA\\
  $^{45}$National Astronomical Observatories, Chinese Academy of Sciences, A20 Datun Road, Chaoyang District, Beijing, 100101, P.~R.~China}
\thanks{$^*$E-mail: \href{zh52@st-andrews.ac.uk}{zh52@st-andrews.ac.uk}}

\begin{abstract}
We present a statistical study of the cool circumgalactic medium (CGM) around massive post-starburst galaxies, traced by Mg\,{\sc ii}\,$\lambda\lambda2796,2803$ absorption in background quasar spectra from the Dark Energy Spectroscopic Instrument (DESI) Data Release~2. Cross-matching $\sim$1.98 million quasar sightlines with luminous red galaxies at $0.4 < z < 0.76$ with stellar masses between $10.6 < \log_{10}(M_*/\mathrm{M}_\odot) < 12$ yields $\sim$143 million galaxy--quasar pairs, 50--60 times more per impact-parameter bin than our previous SDSS-based analysis. We stack the Mg\,{\sc ii} absorption as a function of projected separation and classify the foreground galaxies using mean-field independent component analysis (MFICA), further subdividing the post-starburst population into sub-samples ordered by time since the starburst. The mean Mg\,{\sc ii} equivalent width declines with impact parameter for every class. Post-starburst galaxies show the strongest absorption of any spectral class, reaching $W_{\rm MgII} = 1.58 \pm 0.32$~\AA\ within 71~kpc, well above star-forming ($0.69 \pm 0.15$~\AA) and quiescent ($0.22 \pm 0.05$~\AA) galaxies of comparable stellar mass at the same projected separation. Within the post-starburst phase the absorption profile evolves systematically: the youngest systems show the highest, most centrally concentrated absorption and the steepest radial decline, whereas progressively older systems show weaker inner absorption and enhanced absorption at intermediate radii ($\sim$300--1000~kpc). Independent \textsc{Bagpipes} star-formation-history fitting confirms that this ordering corresponds to an ageing sequence. We interpret this evolution as the outward redistribution of cool gas expelled by feedback-driven winds, consistent with characteristic outflow velocities of a few hundred $\mathrm{km\,s^{-1}}$, with a residual inner cool-gas reservoir persisting throughout the post-starburst phase.

\end{abstract}


\maketitle 



\section{Introduction}

Understanding how galaxies regulate and ultimately cease their star formation remains one of the central challenges of galaxy evolution. Since the peak of cosmic star formation activity at $1<z<3$ \citep{Madau2014}, an increasing fraction of galaxies have transitioned from actively star-forming systems to the quiescent population observed in the local Universe. Explaining this transformation requires understanding not only the processes occurring within galaxies themselves, but also the evolution of the gas reservoirs surrounding them.

The circumgalactic medium (CGM) forms the interface between galaxies and the larger-scale environment. It acts both as a reservoir of gas available for future star formation and as the destination of material expelled through stellar and active galactic nucleus feedback \citep[see review by][]{Tumlinson2017,PerouxHowk2020}. The exchange of mass, metals and energy between galaxies and their CGM is now recognised as a fundamental component of the baryon cycle, linking gas accretion, star formation, feedback and quenching.

Direct observations of cool gas beyond galaxy disks remain challenging. While neutral hydrogen can be detected through 21 cm emission in nearby systems, the low surface brightness of this emission limits studies to relatively local and gas-rich galaxies \citep[e.g.][]{Maddox2021,Dutta2022}. Deep integral-field spectroscopy has recently extended emission-line mapping to higher redshift, revealing extended \ion{Mg}{2} and [\ion{O}{2}] emission around individual galaxies and in stacked samples out to a few tens of kpc \citep[e.g.][]{Dutta2023,Pessa2024}. These measurements provide spatially resolved views of the cool circumgalactic medium that single sightlines cannot, but the steep cosmological dimming of surface brightness restricts them to modest samples and to the brighter, inner regions of haloes. Absorption-line spectroscopy of bright background sources provides a more sensitive probe of diffuse gas extending throughout galaxy haloes. In particular, quasar absorption lines allow the detection of cool circumgalactic material over a broad range of redshifts and environments.

Among the available absorption tracers, the Mg\,II $\lambda2796,2803$ doublet is one of the most extensively studied. Owing to its strength and favourable wavelength range, Mg\,II can be observed over a substantial fraction of cosmic history using large optical spectroscopic surveys. With an ionisation potential similar to that of neutral hydrogen, Mg\,II traces cool ($\sim10^4$ K), relatively dense gas that is often associated with Lyman limit systems and self-shielded structures within the CGM \citep[e.g.][]{BergeronStasinska1986,Churchill2000a,Rigby2002,2025qal1.book.....C}. Although Mg\,II equivalent width is not a direct measure of gas mass because the transition is typically saturated, it provides a sensitive probe of the covering fraction, kinematic complexity and spatial distribution of cool circumgalactic gas.

Large observational studies have established that Mg\,II absorption depends strongly on galaxy properties. Both targeted and blind absorber–galaxy studies and statistical stacking analyses show that stronger absorption is associated with higher stellar masses, smaller impact parameters and increased star formation activity \citep[e.g.][]{ChenHW2010a,Nielsen2013,Lan2018,Huang2021,Anand2021,ChenZ_DESIMgII_2025}. These trends suggest a close connection between the cool CGM and the processes that regulate galaxy growth.

A natural interpretation is that stellar feedback plays an important role in shaping the distribution of circumgalactic gas. Simulations predict that episodes of intense star formation drive metal-enriched outflows into galaxy haloes, transporting cool gas to large distances and contributing to a multiphase CGM \citep[e.g.][]{Muratov2015,Nelson2020,FaucherGiguere2023}. However, substantial uncertainties remain regarding how long this gas survives, how far it propagates, and whether it subsequently returns to the galaxy through recycling flows. Observational constraints on these timescales are particularly limited.

Observational evidence for such feedback-driven outflows is now well established. Blueshifted absorption lines and emission-line kinematics reveal large-scale winds in both local and high-redshift starburst galaxies, with outflow velocities ranging from a few hundred to over $1000\,{\rm km\,s^{-1}}$ in the most extreme systems \citep[e.g.][]{Heckman2000,Rubin2014}. Post-starburst galaxies also exhibit evidence for powerful outflows. \citet{tremontiOriginPoststarburstGalaxies2007} identified Mg\,II absorption signatures indicating cool gas moving at velocities up to $\sim1000\,{\rm km\,s^{-1}}$, demonstrating that substantial quantities of material can be accelerated during or shortly after the starburst phase. Such winds are capable of transporting metals and cool gas far beyond the stellar body of the galaxy, potentially altering the structure of the circumgalactic medium on timescales of hundreds of Myr. However, the ultimate fate of this material remains uncertain. Outflowing gas may escape the halo entirely, remain suspended in the CGM, fragment into smaller clouds, mix with the ambient medium, or eventually return to the galaxy through galactic fountain flows. Determining which of these processes dominates is crucial for understanding how feedback regulates galaxy evolution and the long-term evolution of circumgalactic gas reservoirs.

Post-starburst galaxies provide a unique opportunity to study these processes. These systems have experienced a significant burst of star formation within the past $\sim1$ Gyr followed by rapid quenching, leaving distinctive stellar population signatures that preserve information about the timing and strength of the burst \citep[e.g.][]{Wild2009,Pawlik2018}. Unlike normal star-forming or quiescent galaxies, post-starburst systems therefore offer a means of observing galaxies at different stages following a major feedback event.

Previous studies have already shown that post-starburst galaxies possess unusual circumgalactic gas reservoirs. Ultraviolet absorption-line observations reveal large covering fractions of neutral hydrogen and metal absorption around low-redshift post-starburst galaxies \citep{Borthakur2013,Heckman2017}, while statistical Mg\,II studies demonstrate enhanced cool-gas absorption relative to other galaxy populations. These results indicate that significant quantities of cool gas remain in galaxy haloes after star formation has ceased. However, it remains unclear how this gas evolves following the burst itself.

The most direct precursor to the present work is our own recent statistical study of the cool circumgalactic medium around post-starburst galaxies using SDSS \citep{Harvey2025}. The analysis measured the mean Mg\,{\sc ii} equivalent-width profile as a function of impact parameter for post-starburst galaxies alongside a set of comparison populations, and established that post-starburst systems exhibit systematically stronger cool gas absorption than star-forming or quiescent galaxies of comparable stellar mass, out to projected separations of order of a megaparsec. This provided confirmation, across a large galaxy sample rather than from individual sightlines, that the enhanced cool-gas reservoirs previously inferred from targeted ultraviolet studies \citep{Borthakur2013, Heckman2017} are a generic property of the post-starburst population. The modest numbers of pairs in the SDSS sample nevertheless limited that analysis in two respects that motivate the present study: the green-valley and starburst classes could not be cleanly isolated as distinct populations - the green-valley class blends into the dominant quiescent classification, while the starburst was too rare to yield a usable stack - and the sample did not contain enough galaxy-quasar pairs to subdivide the post-starburst population itself and so track how the cool circumgalactic medium evolves through the post-starburst phase. These are precisely the questions that the substantially larger DESI DR2 sample allows.

In particular, two key questions remain unanswered. First, how does the spatial distribution of Mg\,II-traced gas evolve with time since the starburst event? If starburst-driven outflows redistribute cool gas throughout the halo, one might expect the absorption profile to evolve systematically as the expelled material propagates away from the galaxy. Second, does the strength of the starburst determine the subsequent properties of the CGM? Stronger bursts are expected to inject larger amounts of energy and momentum into the surrounding medium, potentially producing more extended or longer-lived cool-gas reservoirs.

In this paper, we address these questions using Mg\,II absorption measurements around large samples of starburst and post-starburst galaxies. By combining background quasar sightlines with stellar population modelling that constrains both the age and strength of recent bursts, we investigate how the cool circumgalactic medium evolves following a major episode of star formation. We test whether the observed trends are consistent with simple outflow-driven scenarios and explore the implications for feedback, gas recycling and the baryon cycle in massive galaxies.

Throughout this paper we assume a flat $\Lambda$CDM cosmology with $H_0 = 68\,{\rm km\,s^{-1}\,Mpc^{-1}}$, $\Omega_m = 0.3$ and $\Omega_\Lambda = 0.7$.

\section{Data}\label{sec:data}

All data used in this study are from the Dark Energy Spectroscopic Instrument \citep[DESI;][]{DESI2016b, DESI2022}, a robotic, fibre-fed, highly multiplexed spectrograph mounted on the 4-m Mayall telescope at Kitt Peak National Observatory in Arizona. DESI obtains simultaneous spectra of almost 5000 targets over a $\sim 3.2^{\circ}$ field of view \citep{DESI2016b, Silber2023, Miller2024, Poppett2024}, covering the wavelength range 3600--9824~\AA\ at spectral resolution $R \sim 2000$--5500. Designed to measure the expansion history of the Universe through baryon acoustic oscillations and redshift-space distortions \citep{DESI2016a,DESI2024VII}, DESI began main survey operations in 2021 \citep{Schlafly2023} and has since obtained optical spectroscopy for tens of millions of objects, making it the largest spectroscopic survey undertaken to date. Targets are selected from the DESI Legacy Imaging Surveys, and the spectra are reduced with the DESI spectroscopic pipeline \citep{Guy2023}, which processes the raw two-dimensional CCD exposures into calibrated one-dimensional spectra. Object classifications and redshifts are then determined by fitting PCA templates to the reduced spectra with the Redrock code~\citep{Anand2024,Bailey2026}. In this paper we use the spectroscopic quasar and Luminous Red Galaxy (LRG) catalogues from DESI Data Release~2 \citep[DR2;][]{DESI2025II}, whose reduction and release follow the procedures established for DR1 \citep{DESI2026}. We draw both samples directly from the official DESI DR2 quasar and LRG catalogues, and therefore inherit their redshift-quality selection; no additional redshift-quality cuts are applied in this work.

\subsection{The DESI DR2 quasar sample}\label{sec:data_qsos}

The background quasar spectra used in this work are drawn from the DESI DR2 quasar catalogue. The DESI quasar target selection is performed as part of the DESI dark-time survey using photometric selections based on optical $g$, $r$ and $z$ imaging from the DESI Legacy Imaging Surveys combined with mid-infrared \textit{WISE} photometry \citep{Wright2010}. Colour and variability criteria are employed to efficiently distinguish quasars from stars and galaxies while minimising stellar contamination, with the selection optimised for high completeness over the redshift range $0.8 \lesssim z \lesssim 3.5$ \citep{Chaussidon2023}. Redshifts are derived as described above, using the DESI spectroscopic pipeline \citep{Guy2023} and \textsc{Redrock} \citep{Bailey2026,Brodzeller2023}.
 
For this analysis, quasars are restricted to the redshift range $0.8 \lesssim z \lesssim 3$, as shown in Figure \ref{fig:redshifts}, ensuring that the quasar lies securely behind the foreground galaxy sample and that the observed wavelength range used for \mgii\ absorption measurements remains free from contamination by the Lyman-$\alpha$ forest. A further requirement is imposed that each quasar spectrum satisfies a minimum median signal-to-noise ratio of $\mathrm{S/N} > 4$ per pixel measured within a 200~\AA\ window centred on the \mgii\ doublet ($2700$--$2900$~\AA\ in the rest frame of the foreground galaxy). This threshold ensures adequate data quality for continuum estimation and absorption-line measurement in the \mgii\ region, while avoiding the inclusion of low-quality spectra that could bias the equivalent width measurements. No additional quality cuts are applied. After all selection criteria are applied, the final quasar sample contains 1,975,891 unique quasars. 

\subsection{The LRG galaxy sample}\label{sec:data_gals}

\begin{figure*}
	\includegraphics[width=\columnwidth, trim={0.2cm 0 0 0}]{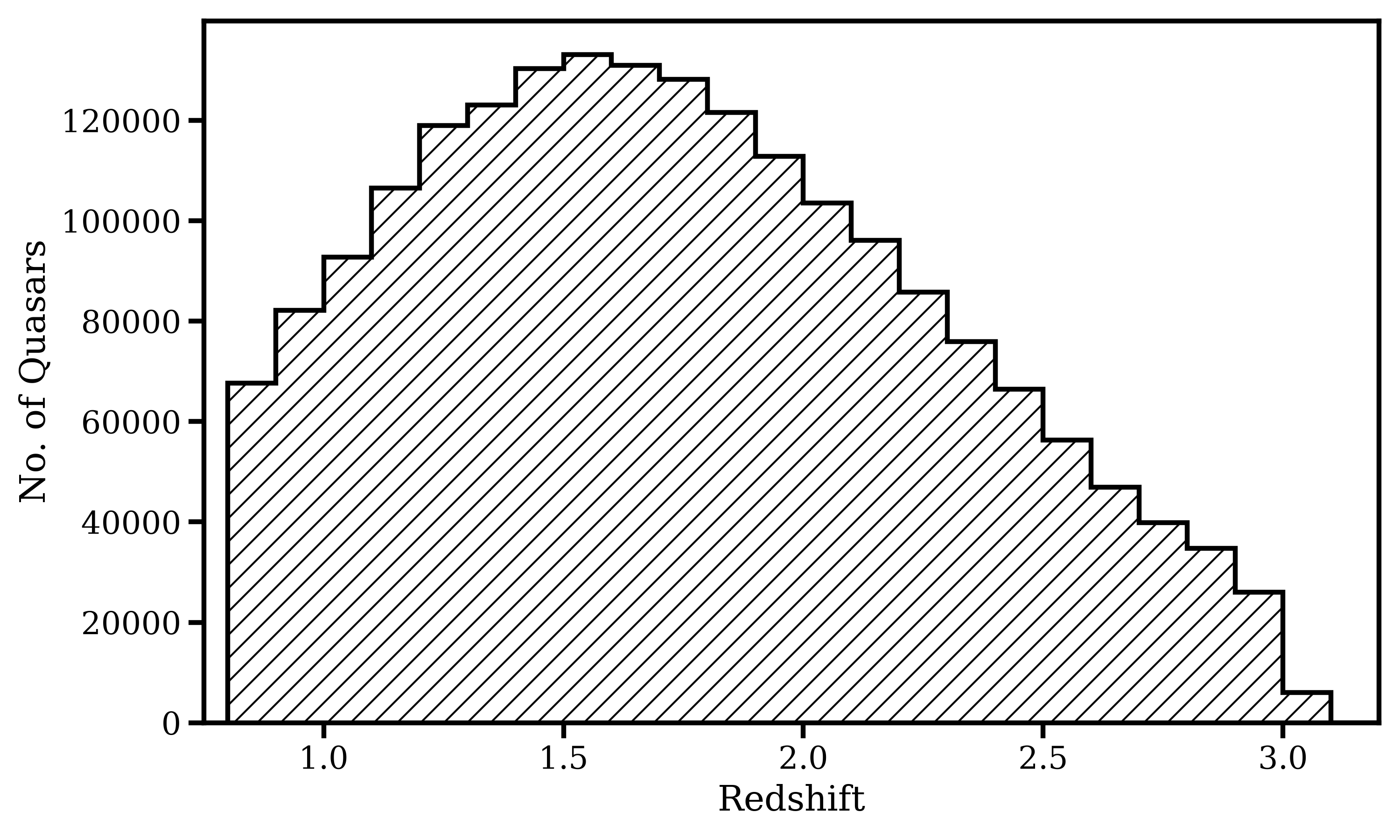}
    \includegraphics[width=\columnwidth, trim={0 0 0 0}]{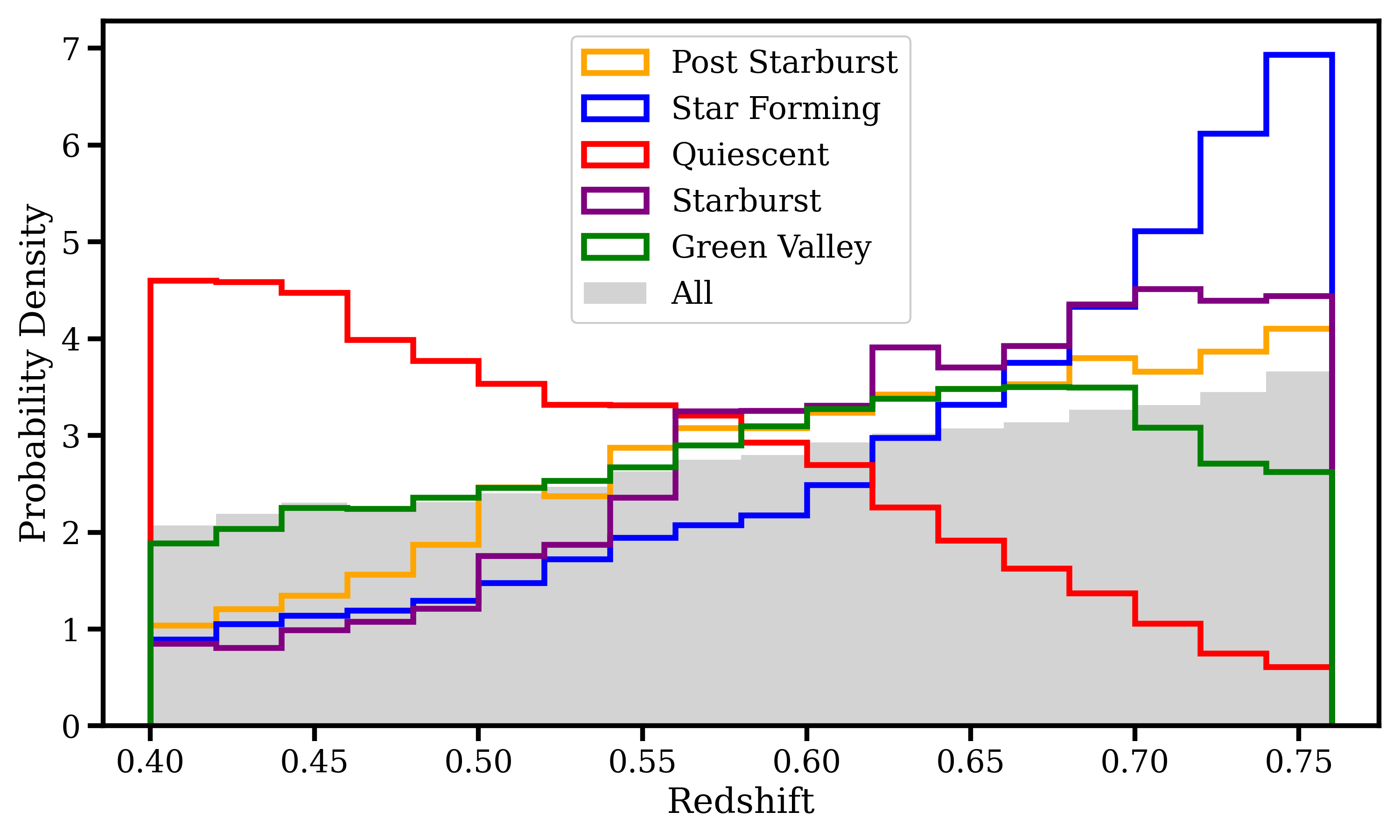}
    \caption{The redshift distribution of the unique DESI DR2 quasars (left panel) and DESI DR2 LRG galaxies (right panel) that make up the galaxy-quasar pairs studied in this paper. The galaxy histograms show the whole sample (grey filled), quiescent (red line), star-forming (blue line), post-starburst (orange line), starburst (purple line) and green valley (green line) sub-samples (see section \ref{sec:data_specclass} for classifications). Each galaxy and each QSO may contribute to multiple pairs in our spectral stacks. }
    \label{fig:redshifts}
\end{figure*}

The foreground galaxy sample is drawn from the luminous red galaxy (LRG) subsample of the DESI DR2 galaxy catalogue. The LRG target selection and its validation are described by \citet{Zhou2023} and also the target-selection pipeline overview of \citet{Myers2023}. Galaxies are selected over the redshift range $0.4 < z < 0.76$, chosen to ensure that the Mg\,\textsc{ii} doublet falls in the observed optical window of the background quasar spectra, to maintain consistency with previous work \citep{Harvey2025} and to match the redshift range over which we apply our spectral classification. This classification uses mean-field independent component analysis (MFICA), which decomposes each galaxy spectrum into components tracing the light from young, intermediate-age and old stellar populations, allowing galaxies to be classified by their recent star-formation history \citep{Krishna2025}; it is described in full in Section~\ref{sec:data_specclass}. The final foreground sample contains 2,239,234 unique luminous red galaxies, with a redshift distribution peaking near $z \approx 0.5$ as shown in Fig.~\ref{fig:redshifts}.
 
Stellar masses for the DESI galaxy sample are obtained from the Stellar Mass and Emission Line value-added catalogue \citep{Zou2024}, in which physical properties are derived by stellar population synthesis fitting with the \textsc{cigale} code \citep{Boquien2019}. The fits combine broad-band $grz$ photometry from the DESI Legacy Imaging Surveys with mid-infrared $W1$ and $W2$ photometry and spectrophotometry from ten artificial bands constructed by convolving the DESI spectra, so that the spectroscopic information contributes directly to the mass estimate. The catalogue used here is a DR2 internal extension of the same pipeline. The stellar mass distribution of the sample is approximately log-normal, peaking at $\log_{10}(M_*/\mathrm{M}_\odot) \approx 11.2$--$11.4$ (Fig.~\ref{fig:galaxy_properties}).

\subsubsection{Galaxy Classifications}\label{sec:data_specclass}

\begin{figure*}
	\includegraphics[width=\columnwidth, trim={0.1 0 0.2cm 0}]{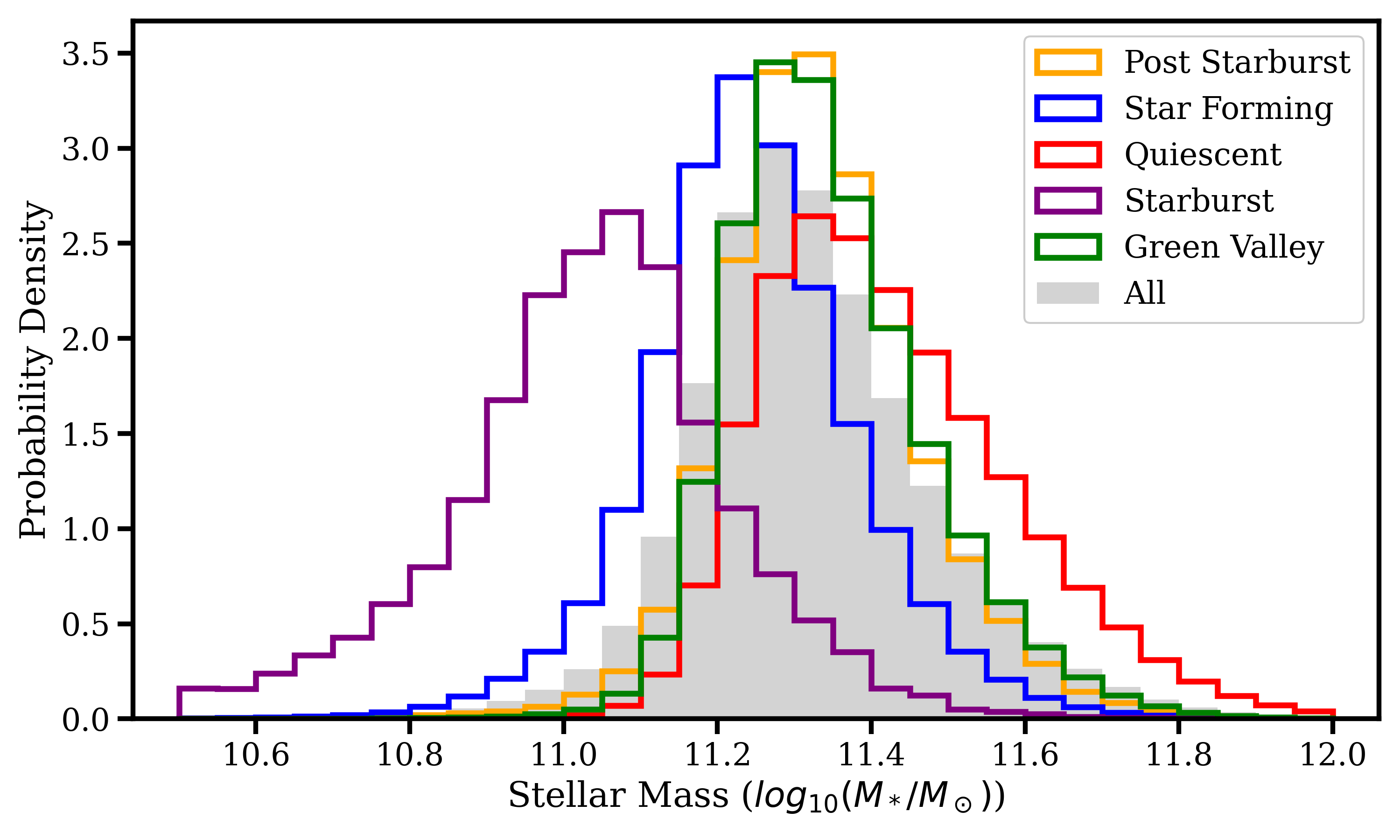}
    \includegraphics[width=\columnwidth, trim={0 0 0.2cm 0}]{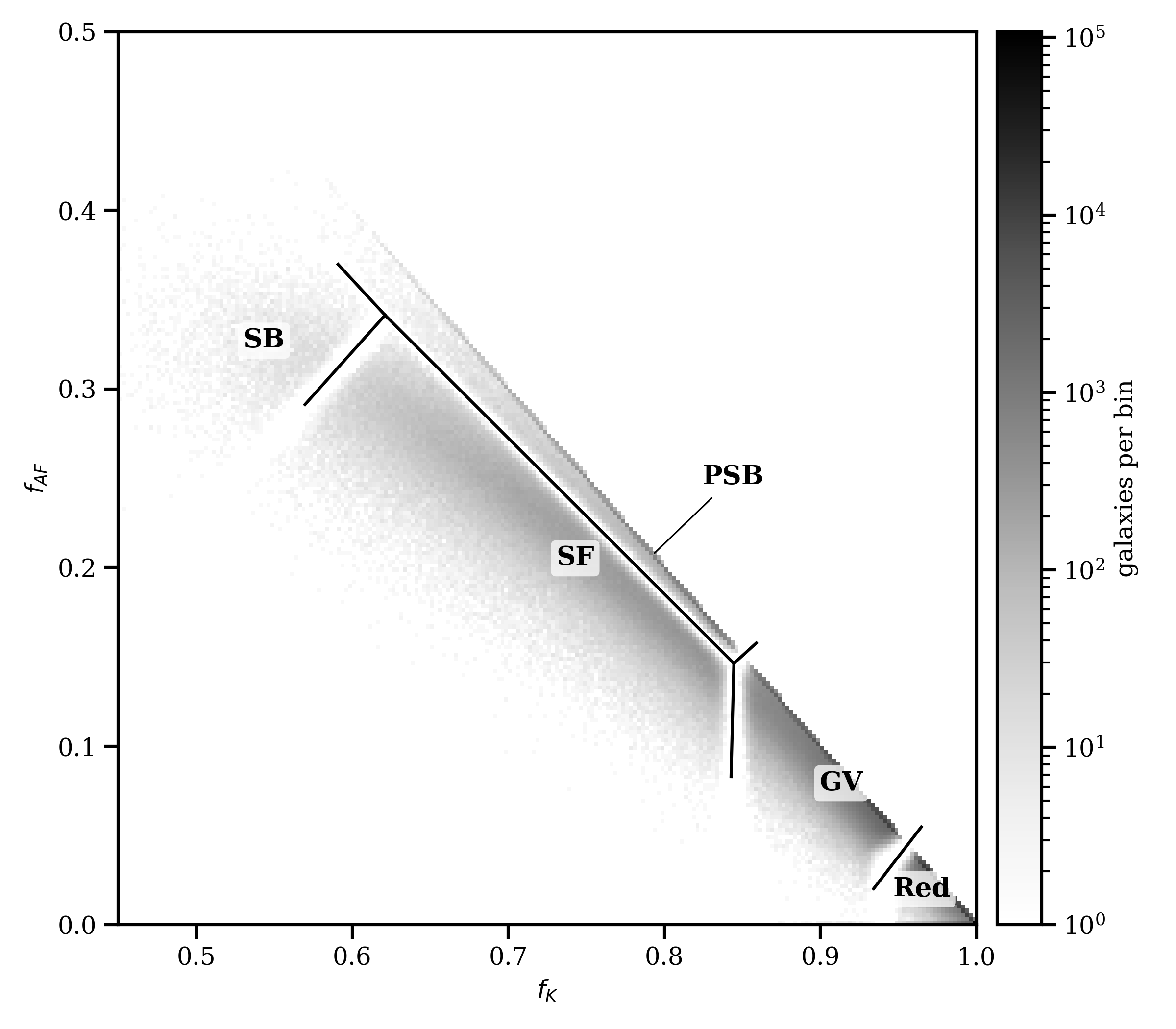}
    \caption{\emph{Left:} The stellar mass distribution of the unique DESI DR2 LRG galaxies that are part of the galaxy-quasar pairs studied in this paper. The histograms show the whole sample (grey filled), quiescent (red line), star-forming (blue line), post-starburst (orange line), starburst (purple line) and green valley (green line) sub-samples.  \emph{Right:} The MFICA derived K and AF component fractions of the galaxy spectra. The K  component fraction represents the fraction of galaxy light between 3300--5200\,\AA\ contributed by long-lived, low-mass stars. The AF component fraction represents the fraction of light contributed by intermediate mass A and F stars. Based on these two light fractions, the galaxy is classified as star forming (``SF''), quiescent (``Q''), post starburst (``PSB''), starburst (''SB'') and green valley (''GV'') as indicated. }
    \label{fig:galaxy_properties}
\end{figure*}

In order to separate galaxies based on their recent star formation history, we fit the galaxy spectra using the mean-field independent component analysis (MFICA) galaxy templates of \citet{Krishna2025}. These seven templates are based on a training sample of SDSS galaxy spectra. The decomposition uses the rest-frame wavelength range $3300$--$5200$~\AA. The first three components predominantly reconstruct the stellar continuum of the galaxy spectra, representing the light contribution of stellar populations of different ages: young (``OB''), intermediate (``AF'') and old (``K''). These three components are used as "fractions", such as the fraction of young, intermediate or old light ($f_{OB}, f_{AF}$ and $f_K$ respectively). A further three components reproduce the nebular emission line strengths, and a final component adjusts the width of the emission lines. The plot of $f_K$ against $f_{AF}$, which is used to determine the classification, is shown in the right panel of Figure \ref{fig:galaxy_properties}.
 
The relative light fractions of the three stellar continuum components provide a data-driven representation of the recent star formation history of each galaxy. Quiescent galaxies are dominated by light from long-lived, low-mass stars and have high K component fractions. Star-forming galaxies are made up of varying combinations of all three components, with starburst galaxies showing the largest OB component fraction due to their ongoing burst of star formation. Post-starburst galaxies, which have recently experienced a starburst that has rapidly quenched, are identified by the strong contribution of intermediate-lifetime A and F stars to their light, giving them a higher AF component fraction than normal star-forming galaxies at fixed K component fraction. All classifications are confirmed by stacking the galaxy spectra for each galaxy class, shown in Appendix \ref{app:stacks}.
 
We divide the sample into five galaxy classes (quiescent, green valley, star-forming, post-starburst and starburst) according to their location in the $(f_K, f_{AF})$ plane, following the classification scheme of \citet{Krishna2025}. Quiescent galaxies are those whose light is almost entirely from old stars ($f_K \gtrsim 0.97$, $f_{AF} \lesssim 0.03$), and green-valley galaxies have small but non-negligible contributions from younger populations ($f_K \approx 0.87$--$0.97$, $f_{AF} \lesssim 0.13$). Star-forming and starburst galaxies have progressively lower $f_K$, with starbursts occupying the lowest-$f_K$ region ($f_K \lesssim 0.63$). Post-starburst galaxies overlap star-forming galaxies in $f_K$ but are separated from them by a higher AF fraction at fixed $f_K$, reflecting the excess light from A and F stars formed in a recent, truncated burst. Because the class regions are defined in the two-dimensional plane, the ranges overlap in each individual fraction. These boundaries follow the physical interpretation of the components described above and match those used in our SDSS analysis \citep{Harvey2025}. We also further sub-divide the post-starburst population into three equal sub-samples ordered by their position in MFICA component space relative to the starburst and quiescent regions. Galaxies located closer to the starburst region are interpreted as having been more recently quenched, while those closer to the quiescent region correspond to more evolved post-starburst systems. This subdivision provides a proxy for time since the starburst event, enabling us to track the evolution of circumgalactic gas properties within the post-starburst phase.
 
To ensure the validity of the objects, we apply a classification confidence threshold based on Monte Carlo resampling. For each galaxy we generate $N =$~1000 realisations of its MFICA component weights by perturbing them within their fitting uncertainties, and re-derive the spectral classification for each realisation. Only galaxies assigned to the same class in at least 90~per~cent of realisations are retained, which improves the separation between classes at the cost of a modest reduction in sample size.

\subsection{Galaxy-quasar pairs}\label{sec:data_pairs}

The galaxy and quasar samples are cross-matched to form line-of-sight pairs by calculating the angular separation between each galaxy and quasar using right ascension and declination. The angular separation is converted to a transverse physical distance (kpc) using the angular diameter distance at the galaxy's redshift. This separation is referred to as the impact parameter. Pairs are initially identified out to 12.8~Mpc, the maximum separation used by \citet{PerezRafols2015}, for consistency with that work; our own analysis uses only pairs within 9.05~Mpc, the outer edge of our largest impact-parameter bin (Table~\ref{tab:pairs}).
 
Galaxy--quasar pairs are removed if any of the following conditions is met:
\begin{itemize}
  \item the quasar redshift is lower than the galaxy redshift;
  \item the galaxy--quasar pair has a velocity difference such that the quasar redshift is within $0.04\,c$ of the galaxy redshift, to exclude absorption line systems associated with quasar outflows;
  \item the galaxy's \mgii\ absorption feature falls within the Lyman-$\alpha$ forest region of the quasar spectrum (i.e. $< 1215.67$~\AA\ in the quasar rest frame);
  \item the galaxy's \mgii\ absorption feature falls within strong emission line regions of the quasar spectrum (specifically within 30~\AA\ centred on C\,\textsc{iv}~$\lambda 1549$ and 20~\AA\ centred on C\,\textsc{iii]}~$\lambda 1909$).
\end{itemize}
 
This results in a total of approximately 143 million galaxy-quasar pairs drawn from 1,975,891 unique quasars. As shown in Table~\ref{tab:pairs}, the DESI DR2 sample provides approximately 50–60 times more galaxy–quasar pairs per impact-parameter bin than the previous SDSS-based analysis of \citet{Harvey2025}, reflecting the substantially larger sky coverage and increased spectroscopic number density of the DESI survey.
 
\begin{table*}
  \centering
  \caption{The number of galaxy - quasar pairs in bins of impact parameter for the DESI DR2 sample used in this work, alongside the SDSS DR7/DR12 pair counts from \citet{Harvey2025} for comparison. Mean impact parameters are calculated from the DESI DR2 pairs in each bin, where the bin edges match those of previous work and the means differ only marginally between the two samples.}
  \begin{tabular}{c c c c}
    \hline
    Impact Parameter & Mean Impact & \multicolumn{2}{c}{Number of Pairs} \\
    (kpc)            & Parameter (kpc) & SDSS DR7/DR12 & DESI DR2 \\
    \hline
     0  -  71          &   51   &        115   &        7\,277   \\
     71  -  141         &  112   &        420   &       23\,319   \\
     141  -  282        &  219   &      1\,854   &      102\,070   \\
     282  -  566        &  442   &      7\,577   &      478\,445   \\
     566  -  1131       &  883   &     30\,260   &    2\,107\,493  \\
     1131  -  2262      & 1\,757 &    122\,451   &    8\,564\,566  \\
     2262  -  4525      & 3\,421 &    488\,348   &   34\,021\,270  \\
     4525  -  9051      & 7\,800 &  1\,947\,357  &   97\,297\,198  \\
    \hline
  \end{tabular}
  \label{tab:pairs}
\end{table*}

We do not mask the quasar Ly$\alpha$ emission line, nor do we exclude broad absorption line (BAL) quasars. Continuum residuals around strong emission lines could in principle bias the measured absorption, since the masked region is always centred on the foreground Mg\,\textsc{ii} wavelength. However, the redshift-randomisation test (Section~\ref{sec:nulltest}), which repeats the masking and continuum estimation at the shuffled redshifts, recovers no absorption at any impact parameter. BAL troughs, being uncorrelated with the foreground galaxy redshift, add only noise to the stacks.

\section{Method}\label{sec:methods}
In this section we detail the procedures used to remove the quasar continuum from the background quasar spectra and stack the spectra at the redshift of the foreground galaxies to create high signal-to-noise stacked spectra of the \mgii\ absorption line. We then describe the measurement of the galaxy rest-frame equivalent width of the \mgii\ absorption doublet ($W_{\rm MgII}$).
 
\subsection{Removing the quasar continuum}
 
To remove the quasar emission continuum, we apply a series of median filters and sigma-clipping steps to each background quasar spectrum, following the procedure described in \citet{Harvey2025}. Before applying the filtering, we mask a 20~\AA\ wide region centred on the \mgii\ doublet in the galaxy rest frame ($\pm10\text{\AA}$) to prevent the signal of interest from being absorbed into the continuum estimate. Strong observed-frame sky emission lines are also masked. We apply a running median filter of 45 pixels, including across the masked regions, and divide the quasar flux and error arrays by this running median. Pixels with absolute flux/error $> 2.5$ are iteratively masked and replaced by the median filter value. Two final iterations with filter widths of 15 and 65 pixels are applied to reduce any residual continuum curvature near the \mgii\ doublet.
 
\subsection{Stacking}

To create stacked spectra, individual continuum-divided spectra are shifted to the absorber rest frame by dividing their observed wavelengths by $(1+z_{\rm gal})$, and the region between $2750$--$2850$~\AA\ is extracted. Rather than interpolating or resampling the flux onto a new grid, the shifted pixels are accumulated onto a common rest-frame wavelength grid, with each native pixel assigned to the bin into which it fell; this is what we refer to as stacking without rebinning. A weighted mean stack is formed in each impact-parameter bin. Each continuum-normalised spectrum is weighted by the inverse of its variance, $w_i = 1/\sigma_i^2$, where $\sigma_i^2$ is the pipeline flux variance averaged over a 100~\AA\ window centred on the \mgii\ doublet. Noisier spectra therefore contribute proportionally less to the stack. The stacked flux is normalised by dividing by the total of the weights contributing at each wavelength, and the pixel variances are propagated through the same weighted accumulation so that flux and variance are treated consistently.
 
\subsection{Equivalent widths}
 
The strength of the \mgii\ doublet is measured by directly integrating the flux deficit over the region $2791$--$2808$~\AA, encompassing both doublet lines. No Voigt-profile modelling is performed. Uncertainties are estimated by bootstrap resampling the galaxy--quasar pairs with replacement 1000 times, repeating the full stacking and integration procedure, and taking the standard deviation of the resulting distribution.

\subsection{Validation with randomised redshifts}
\label{sec:nulltest}
 
To verify that our results are not driven by noise or systematic effects in the stacking procedure, we randomise the galaxy redshifts by reassigning each galaxy a redshift drawn from another galaxy in the sample, preserving the overall redshift distribution. Repeating the full procedure with these shuffled redshifts, including the Mg\,\textsc{ii} masking and continuum estimation, removes any physical association between galaxies and absorbers while preserving any systematic arising from the analysis itself. No coherent \mgii\ absorption feature is detected in the shuffled stacks at any impact parameter, with measured equivalent widths consistent with zero across the full radial range probed ($0$--$9.05$~Mpc), confirming that our measured signals are physically associated with the foreground galaxy population and are not driven by noise, systematic continuum structure, or artefacts of the stacking method. The scatter of the shuffled-redshift equivalent widths provides an estimate of the noise floor of our stacks - in every impact-parameter
bin this scatter is $\lesssim 0.005$\,\AA, below the equivalent widths measured from the true stacks.

\section{Results}\label{sec:results}

\subsection{Global \mgii\ absorption profile}
\label{sec:results_global}
 
\begin{figure*}
  \centering
  \includegraphics[width=\textwidth]{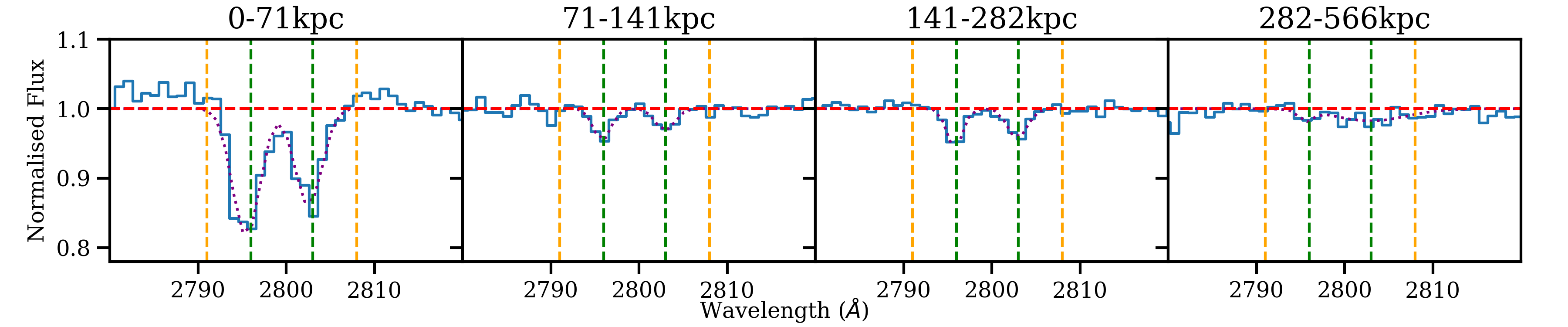}
  \includegraphics[width=\textwidth]{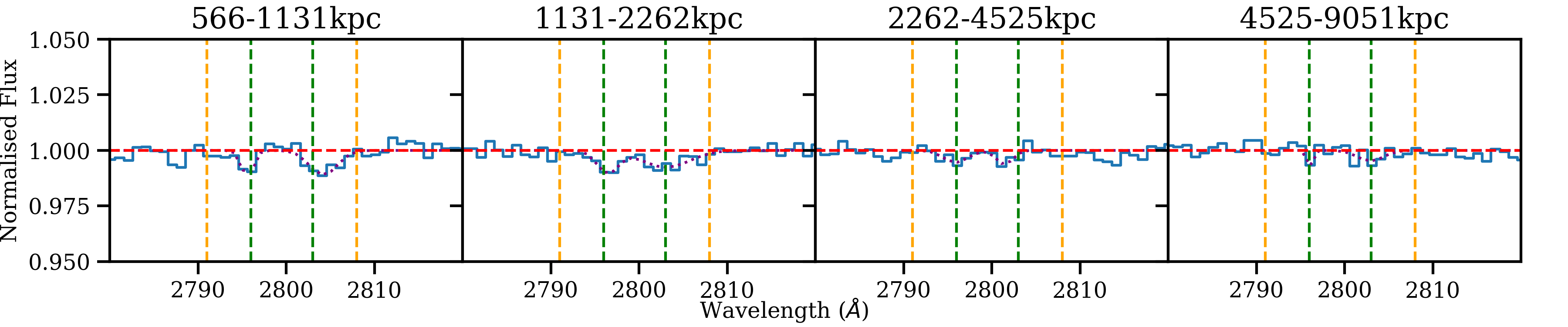}
  \caption{Stacked spectra around the \mgii\ absorption doublet in the rest frame of the galaxies for the inner four (top) and outer four (bottom) impact parameter bins. The green dashed lines show the expected positions of the \mgii\ doublet, and the orange dashed lines show the wavelengths over which the equivalent width is measured. A double Gaussian profile fit is overlaid for visualisation (purple dotted lines). An absorption signature is   detected at all impact parameters.}
  \label{fig:allstacks}
\end{figure*}

\begin{figure*}
  \centering
  \includegraphics[width=\textwidth]{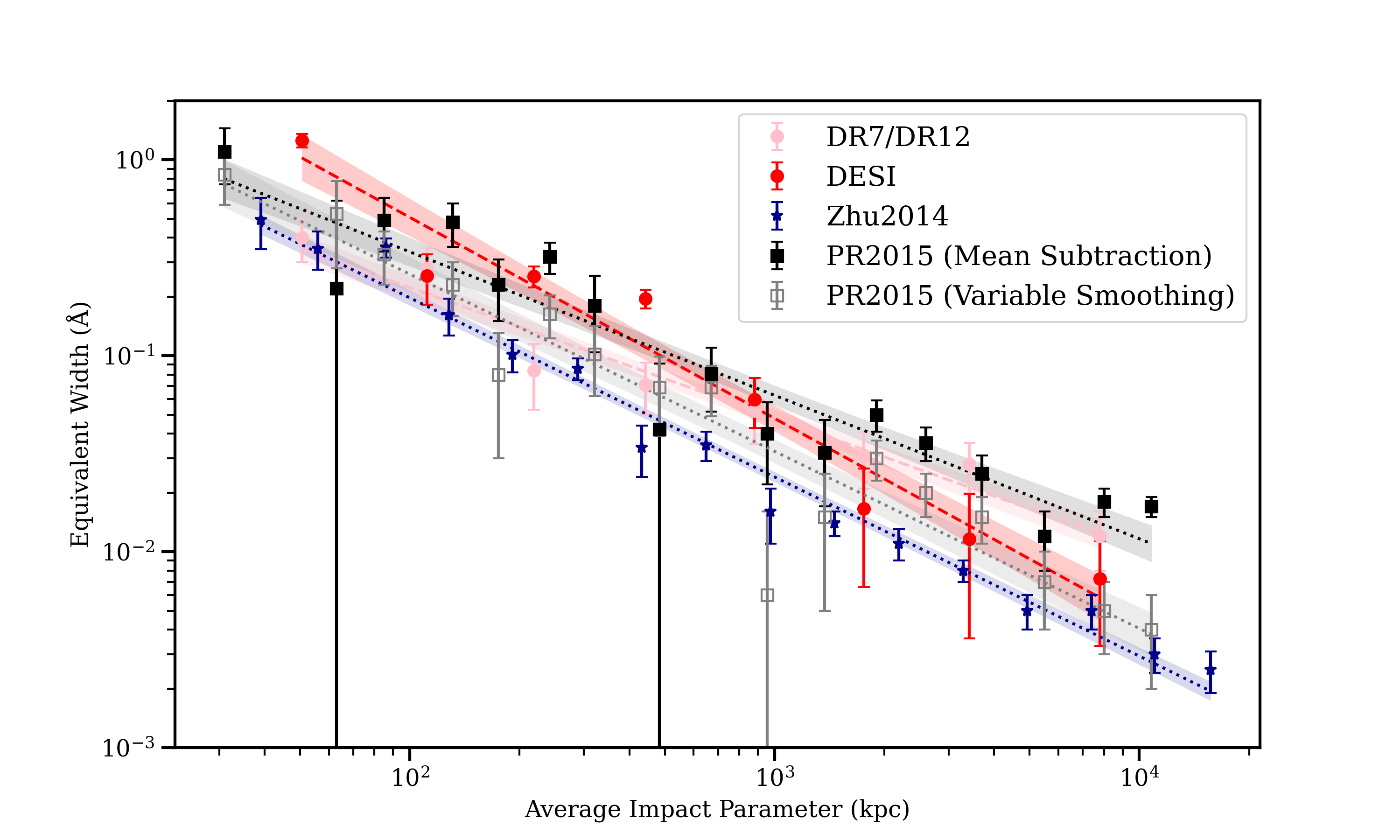}
  \caption{The equivalent width of the \mgii\ absorption doublet as a function of average impact parameter for the full DESI DR2 sample (red points), compared to the SDSS-based result from \citet{Harvey2025} (pink points), \citet{Zhu2014} (blue) and \citet{PerezRafols2015} (black/grey). Dotted lines show least-squares linear fits to each dataset. The DESI results are consistent with all prior measurements while providing improved precision.}
  \label{fig:eqw_b}
\end{figure*}

Fig.~\ref{fig:allstacks} shows the stacked spectra around the \mgii\ doublet for the total galaxy sample across all eight impact-parameter bins. A significant absorption signature is detected from within 71~kpc of the galaxy out to $\sim 1$~Mpc. At larger separations the signal remains above both the detection threshold and the noise floor established by the shuffled-redshift stacks (Section~\ref{sec:nulltest}), though at progressively lower amplitude. As shown in Fig.~\ref{fig:eqw_b}, the equivalent width declines monotonically with increasing impact parameter. This figure also shows the results compared to previous literature.

\begin{table*}[]
    \centering
    \begin{tabular}{c|c|c|c|c|c}
          Impact Parameter (kpc) &  \multicolumn{5}{c}{No. of Pairs} \\
          & Quiescent & Green Valley & Star Forming & Post-Starburst & Starburst \\
         \hline
       0 - 71 & 1357 & 1830 & 2720 & 219 & 29 \\
       71 - 141 & 4880 & 5813 & 8191 & 693 & 87 \\
       141 - 282 & 21175 & 25796 & 36029 & 2903 & 381 \\
       282 - 566 & 99589 & 120340 & 166866 & 13916 & 1831 \\
       566 - 1131 & 440896 & 532772 & 733322 & 61656 & 8067 \\
       1131 - 2262 & 1799075 & 2167940 & 2972338 & 251158 & 32606 \\
       2262 - 4525 & 7144737 & 8611902 & 11799695 & 997080 & 129883 \\
       4525 - 9051 & 10845684 & 12751235 & 19856457 & 1345685 & 256485 \\
    \end{tabular}
    \caption{The number of quasar-galaxy pairs contributing to each bin, as a function of projected impact parameter and MFICA galaxy classification. Impact parameter bins are defined in projected physical separation from the foreground galaxy. These pair counts determine the statistical weight of each stack and illustrate the relative sampling of different galaxy populations.}
    \label{tab:galaxy_pairs}
\end{table*}

\begin{table*}[]
\centering
\begin{tabular}{c|c|c}
Impact Parameter (kpc) & Quiescent & Green Valley \\
\hline
0--71      & $0.2246 \pm 0.0512$ & $0.525 \pm 0.0912$ \\
71--141    & $0.1401 \pm 0.0318$ & $0.193 \pm 0.0302$ \\
141--282   & $0.0802 \pm 0.0225$ & $0.118 \pm 0.0212$ \\
282--566   & $0.0512 \pm 0.0121$ & $0.067 \pm 0.0098$ \\
566--1131  & $0.0287 \pm 0.0092$ & $0.042 \pm 0.0083$ \\
1131--2262 & $0.0254 \pm 0.0081$ & $0.040 \pm 0.0064$ \\
2262--4525 & $0.0214 \pm 0.0043$ & $0.035 \pm 0.0049$ \\
4525--9051 & $0.0215 \pm 0.0028$ & $0.0301 \pm 0.0042$ \\
\end{tabular}
\begin{tabular}{c|c|c|c}
Impact Parameter (kpc) & Star Forming & Post-Starburst & Starburst \\
\hline
0--71      & $0.686 \pm 0.153$ & $1.578 \pm 0.321$ & $2.320 \pm 0.618$ \\
71--141    & $0.341 \pm 0.081$ & $0.757 \pm 0.154$ & $0.166 \pm 0.038$ \\
141--282   & $0.097 \pm 0.028$ & $0.464 \pm 0.079$ & $0.117 \pm 0.033$ \\
282--566   & $0.087 \pm 0.016$ & $0.165 \pm 0.052$ & $0.069 \pm 0.023$ \\
566--1131  & $0.042 \pm 0.011$ & $0.102 \pm 0.029$ & $0.062 \pm 0.009$ \\
1131--2262 & $0.035 \pm 0.007$ & $0.051 \pm 0.010$ & $0.054 \pm 0.008$ \\
2262--4525 & $0.033 \pm 0.006$ & $0.046 \pm 0.007$ & $0.040 \pm 0.007$ \\
4525--9051 & $0.030 \pm 0.005$ & $0.039 \pm 0.005$ & $0.032 \pm 0.006$ \\
\end{tabular}
\caption{Mean Mg\,\textsc{ii} equivalent width (\AA) as a function of
impact parameter for galaxies in the DESI galaxy--quasar pair sample,
separated by MFICA classification: star-forming (SF), starburst (SB),
post-starburst (PSB), green valley (GV), and quiescent (Q).
Measurements are obtained from stacked spectra in each
impact-parameter bin; uncertainties are $1\sigma$ from bootstrap
resampling.}
\label{tab:galclass_ew}
\end{table*}

\subsection{\mgii\ absorption as a function of galaxy spectral class}
\label{sec:results_class}
 
\begin{figure*}
  \centering
  \includegraphics[width=\textwidth]{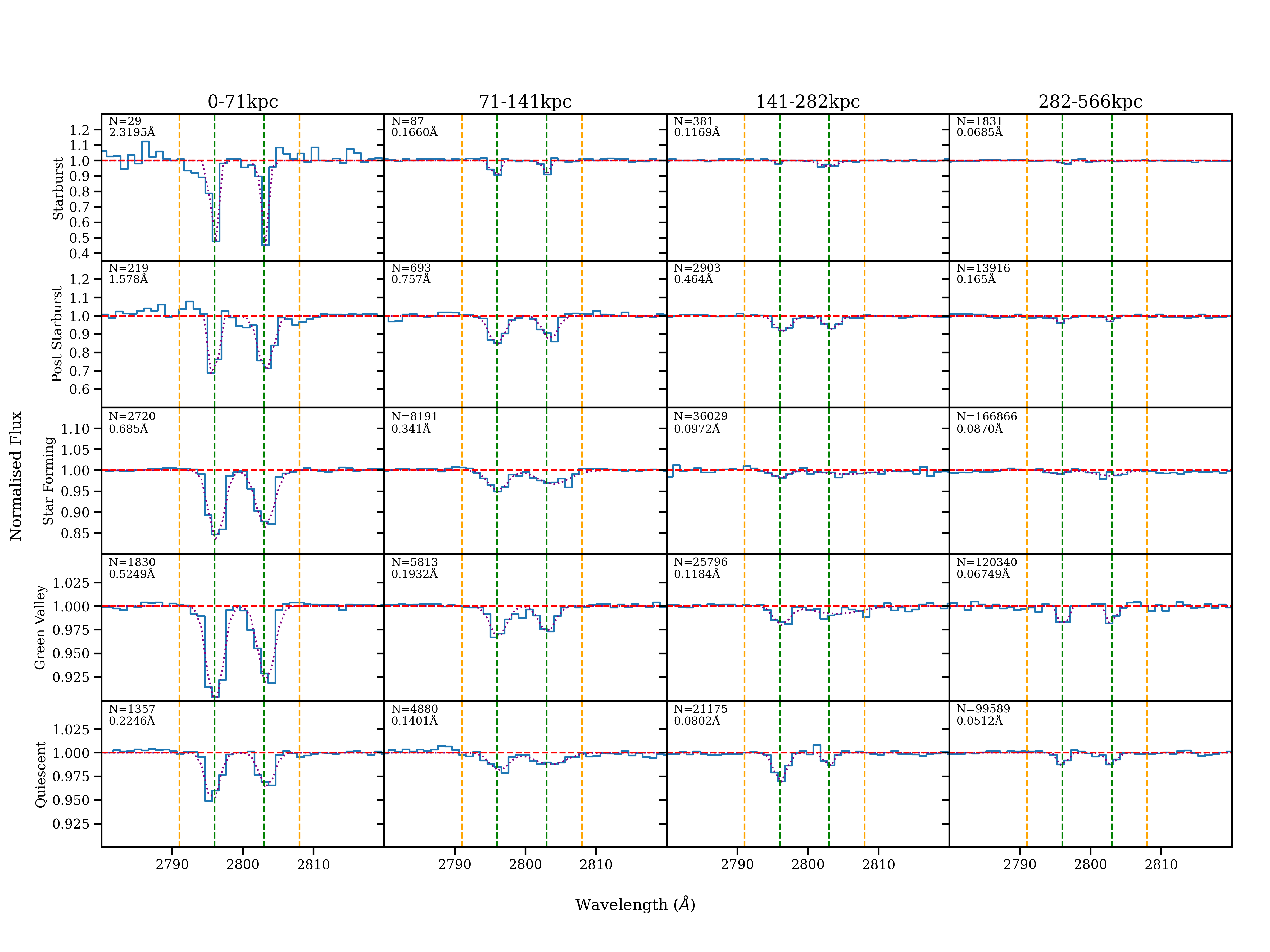}
  \caption{Stacked spectra around the \mgii\ doublet in the rest frame of the galaxies for the inner four impact parameter bins, for each MFICA spectral class: post-starburst (top), star-forming, quiescent, green valley, and starburst (bottom). The absorption signal decreases with impact parameter for all galaxy types. Post-starburst and starburst galaxies show the strongest absorption at small separations; quiescent galaxies show the weakest signal.}
  \label{fig:classstacks}
\end{figure*}

\begin{figure*}
  \centering
  \includegraphics[width=\textwidth]{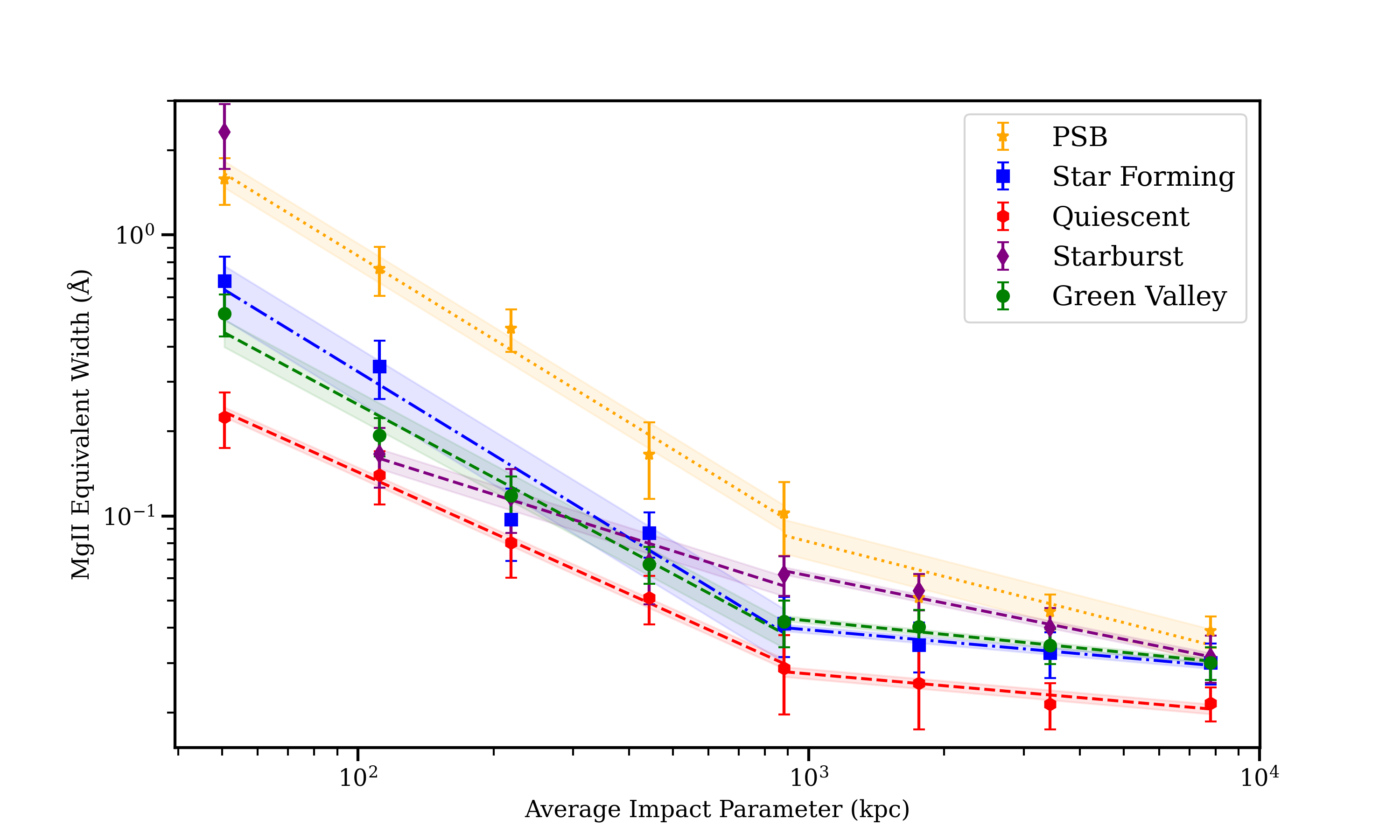}
  \caption{The equivalent width of the \mgii\ absorption doublet as a function of average impact parameter for each MFICA spectral class: post-starburst (orange), star-forming (blue), quiescent (red), green valley (green) and Starburst (purple). Lines indicate least-squares linear fits, independently within and beyond 1~Mpc. Post-starburst and starburst galaxies show the strongest absorption at small separations; all populations converge beyond 1~Mpc.}
  \label{fig:eqw_b_class}
\end{figure*}

Fig.~\ref{fig:classstacks} shows the stacked \mgii\ spectra for the inner four impact parameter bins, separated by galaxy spectral class. The equivalent width profiles for all five classes  -  quiescent, green valley, star-forming, starburst and post-starburst - are presented in Fig.~\ref{fig:eqw_b_class}, with measurements shown in Table~\ref{tab:galclass_ew} and number of pairs in each classification bin in Table \ref{tab:galaxy_pairs}. The galaxy stacks confirming the classification split are shown in Appendix \ref{app:stacks}.
 
All galaxy populations exhibit a declining absorption profile with increasing impact parameter. However, systematic differences between populations are clearly evident. Post-starburst galaxies show the strongest \mgii\ absorption at nearly all impact parameters, with an equivalent width at the innermost bin ($< 71$~kpc) of $1.58 \pm 0.32$~\AA, substantially exceeding that of star-forming ($0.69 \pm 0.15$~\AA) and quiescent ($0.22 \pm 0.05$~\AA) galaxies at the same projected separation. Starburst galaxies show comparably strong absorption ($2.32 \pm 0.62$~\AA) at very small impact parameters, consistent with cool gas being concentrated close to the galaxy during the ongoing burst, but their profile falls off more steeply with radius compared to post-starburst systems, where the absorption remains elevated to larger projected separations.
 
The quiescent and green-valley populations occupy an intermediate regime, with green-valley galaxies lying between the star-forming and quiescent populations as expected for a transitional class. Beyond $\sim 1$~Mpc, the equivalent width profiles of all five sub-samples converge, with differences between populations becoming small compared to the measurement uncertainties.
 
These results from the DESI DR2 dataset are consistent with and extend the findings of \citet{Harvey2025} based on SDSS data. The increased sample size allows us to confirm that the enhanced \mgii\ absorption around post-starburst galaxies is a robust feature of this population, and to include the starburst and green-valley classes that could not be cleanly separated in the smaller SDSS CMASS sample.
 
\subsection{The evolution of \mgii\ absorption across the post-starburst population}
\label{sec:results_psb_evolution}

The analysis presented so far has treated post-starburst galaxies as a single population. In this section we sub-divide them according to the time elapsed since their starburst, and ask whether the \mgii\ absorption profile evolves along that sequence. This is done in two stages. We first establish an ordering of the population by burst age, using the position of each galaxy in the MFICA plane to divide the sample into tertiles, and testing that ordering against direct star formation history fits to a high signal-to-noise subsample (Section~\ref{sec:results_bagpipes}). We then measure the stacked \mgii\ equivalent width profile of each tertile as a function of impact parameter (Section~\ref{sec:results_psb}).

\subsubsection{Determining burst age}
\label{sec:results_bagpipes}

\begin{figure*}
  \centering
  \includegraphics[width=\textwidth]{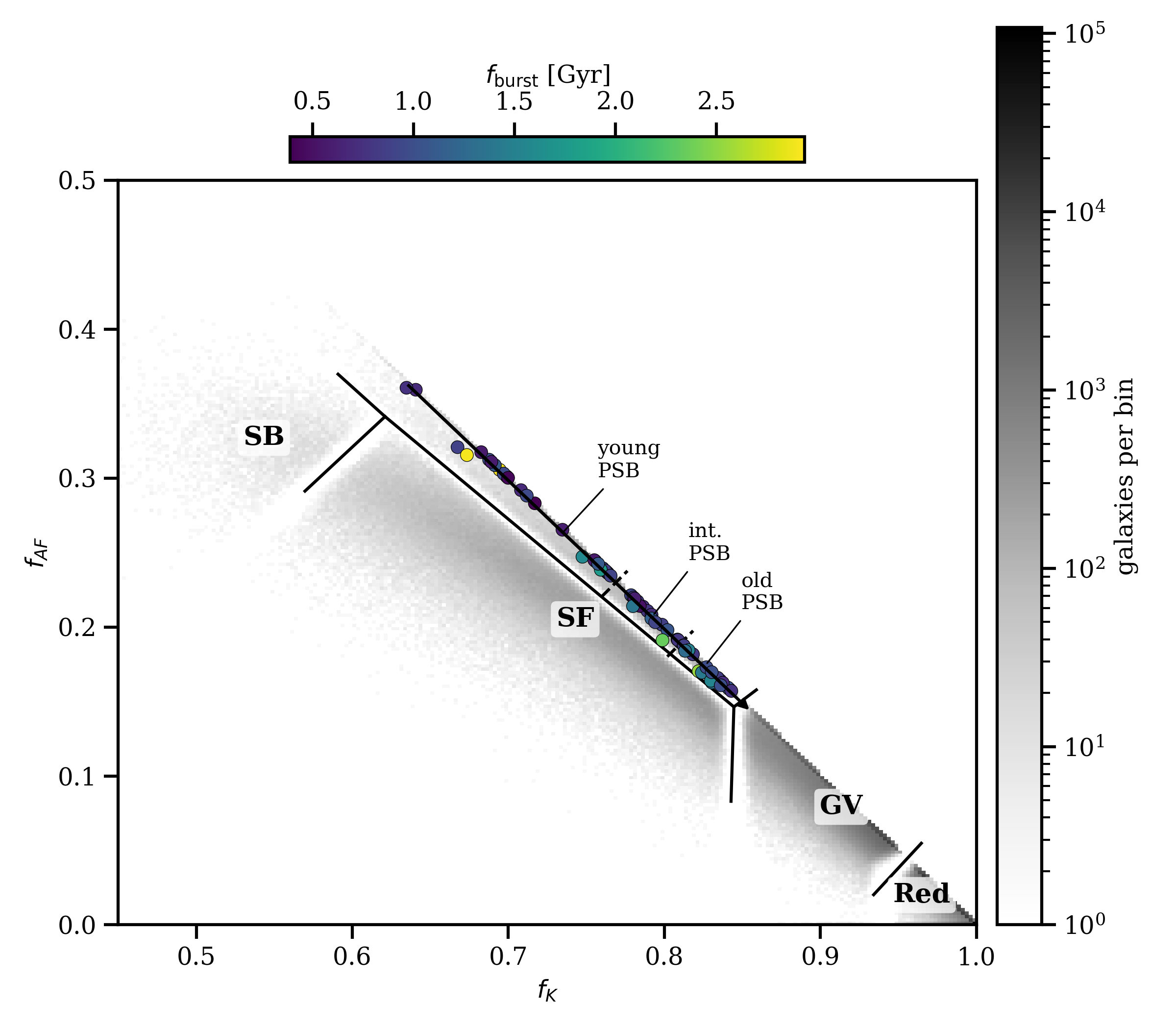}
  \caption{The location of the \textsc{Bagpipes} subsample (circles, coloured by $t_{\rm burst,50}$) within the $(f_K, f_{AF})$ plane of the full classified DESI sample. Background points show the parent classification into starburst (SB), star-forming (SF), quiescent (Red), green valley (GV) and unclassified galaxies, with the post-starburst further divided into three sub-regions following the same scheme adopted in Section~\ref{sec:data_specclass}. The subsample spans the full post-starburst sample used in the main analysis.}
  \label{fig:bagpipes_classification}
\end{figure*}

We approach the ordering of the post-starburst population in two ways. The first uses the spectral classification itself. We parametrise position along the post-starburst locus in the $(f_K, f_{AF})$ plane by a coordinate $u$, with $u=0$ at the starburst end and increasing towards the quiescent end, and assign each galaxy the value of $u$ at its projected location along this axis (Fig.~\ref{fig:bagpipes_classification}). Larger values of $u$ correspond to more $f_K$-dominated, quiescent-like spectra. Ordering the population by $u$ and dividing it into three equal-sized samples gives the youngest, intermediate and oldest sub-samples used throughout the remainder of this section. This sub-division rests on the assumption, introduced in Section~\ref{sec:data_specclass}, that position in the MFICA plane traces the time elapsed since the starburst. In principle, position along the post-starburst locus is degenerate between burst age and burst strength: galaxies at small $u$ may host either a more recent or a stronger burst, since both raise the A/F light fraction relative to the old stellar population, while galaxies at large $u$ may host either an older or a weaker burst. We break this degeneracy with the \textsc{Bagpipes} fits below, which constrain the burst age and burst mass fraction separately.

The second approach tests this assumption independently, by fitting star
formation histories directly. We identify a subsample of 63 post-starburst galaxies drawn from the DESI DR2 catalogue with sufficiently high per-pixel signal-to-noise ($\mathrm{S/N} > 8$) to permit full spectral star formation history (SFH) fitting with the \textsc{Bagpipes} code \citep{Carnall2018}. Fitting the stacks as a whole is not viable: the galaxies contributing to a stack have a range of star formation histories, burst times and burst strengths, so the stacked spectrum is an average that does not correspond to any single SFH and cannot be meaningfully fit. We therefore work with individual spectra, which restricts us to the brightest objects. In interpreting these fits we assume that, within each tertile, the high signal-to-noise galaxies share the burst-age distribution of the full tertile. Because younger or stronger bursts are brighter, this subsample may be biased towards such systems; however, any such bias would affect all three tertiles alike, shifting the absolute ages rather than their ordering along $u$, which is all our test relies on. Fig.~\ref{fig:bagpipes_classification} shows the location of these 63 galaxies within the full classified DESI sample, overlaid on the three
post-starburst sub-divisions.

We adopt the two-component SFH model of \citet{Leung2024}, in which an old, exponentially declining stellar population is combined with a more recent burst, and refer the reader to that work for a full description of the model and its priors. \textsc{Bagpipes} uses the 2016 version of the \citet{BC03} stellar population models with a \citet{Kroupa2001} initial mass function (IMF). This parametrisation is well suited to post-starburst systems, as it recovers the rapid rise and subsequent decline in star formation rate that characterises the population. For each galaxy, the fit provides posterior median estimates of the burst age ($t_{\rm burst,50}$), the fraction of stellar mass formed in the burst ($f_{\rm burst,50}$), the timescale over which star formation subsequently declined ($\tau_{\rm quench,50}$), and the cosmic time at which the galaxy quenched ($t_{\rm quench,50}$). Two galaxies are removed from the subsample because their fits do not converge acceptably, leaving 61 galaxies in the analysis below.

\begin{figure}
  \centering
  \includegraphics[width=\columnwidth]{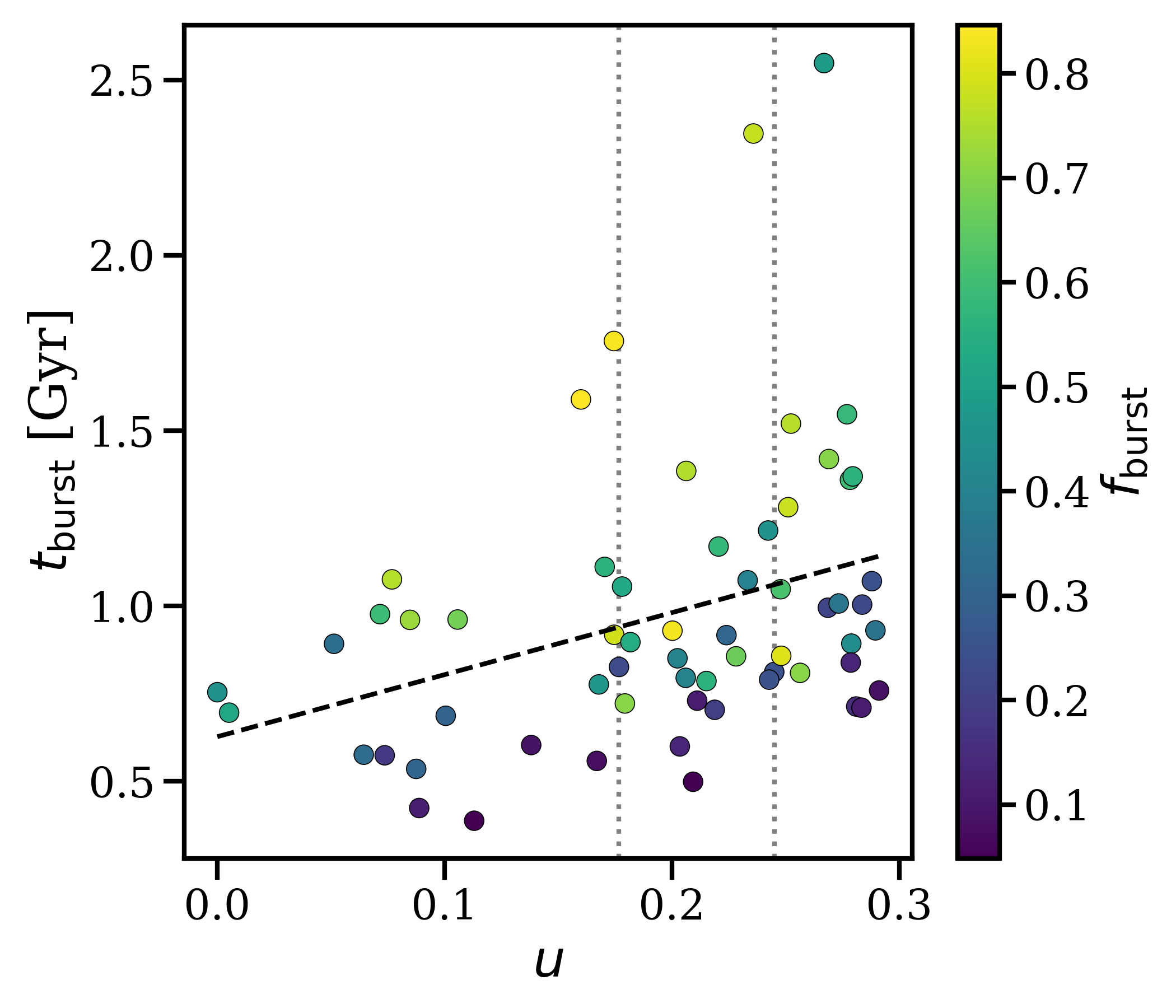}
  \caption{$t_{\rm burst,50}$ as a function of position $u$, with a least-squares linear fit overlaid (dashed line). Here $u$ is the coordinate along the post-starburst locus in the $(f_K, f_{AF})$ plane (Fig.~\ref{fig:bagpipes_classification}), with $u=0$ at the starburst end and increasing towards the quiescent end.}
  \label{fig:bagpipes_age_position}
\end{figure}

Fig.~\ref{fig:bagpipes_age_position} shows $t_{\rm burst,50}$ as a function of position along the $u$ direction. A positive trend is visible: galaxies at larger $u$ tend to host older stellar bursts, in the sense expected if $u$ tracks time since the burst.

\begin{figure}
  \centering
  \includegraphics[width=\columnwidth]{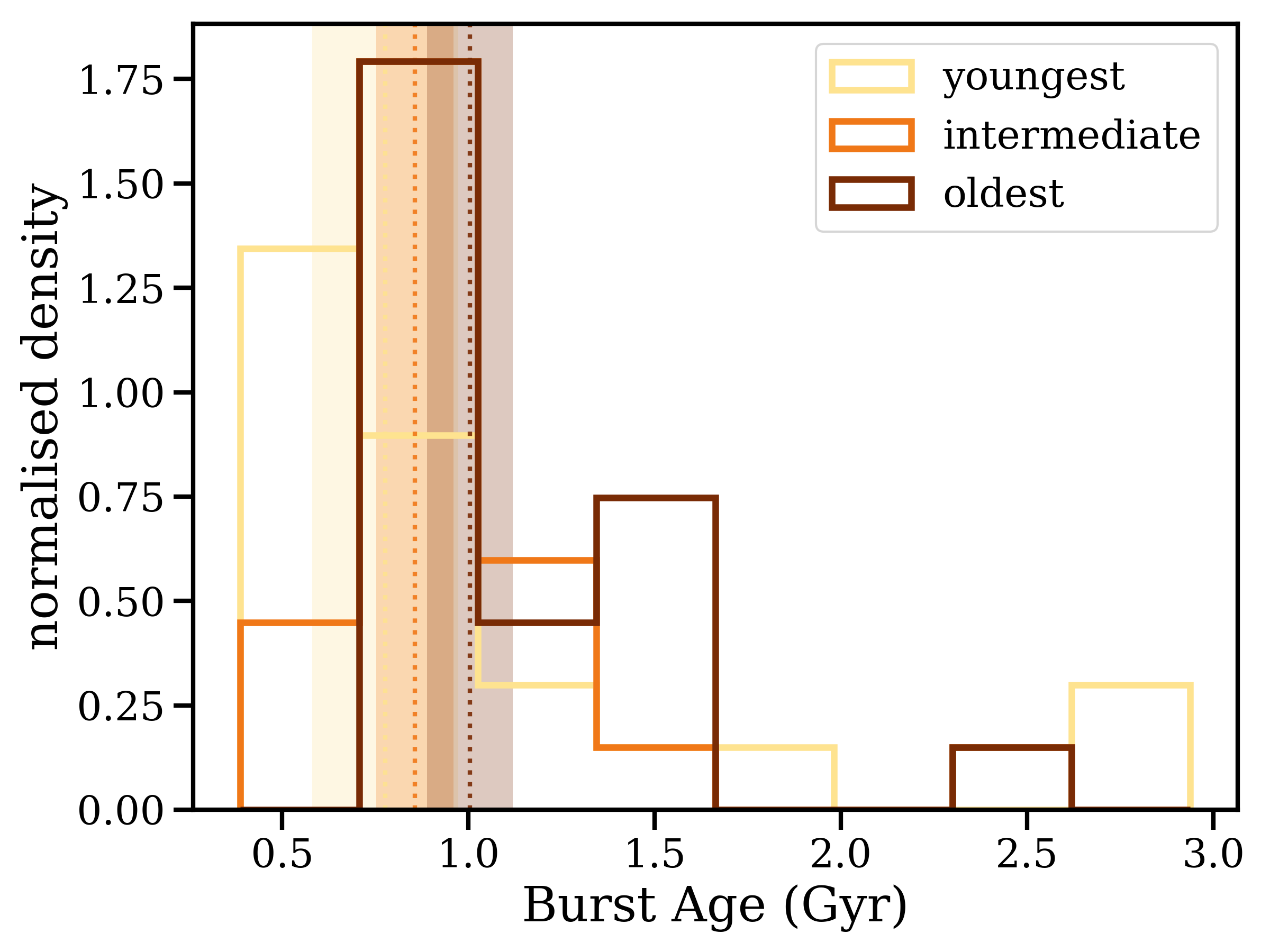}
  \caption{Normalised distributions of $t_{\rm burst,50}$ for the \textsc{Bagpipes} subsample, divided into three equal-sized samples along the $u$ direction (youngest: closest to the starburst end; oldest: closest to the quiescent end), matching the three-fold post-starburst sub-division used in Section~\ref{sec:results_psb}. Dotted vertical lines show the median burst age in each sample, which increases from the youngest to the oldest.}
  \label{fig:bagpipes_age_hist}
\end{figure}

We further test this by dividing the subsample into three equal-sized samples along the $u$ direction, matching the tertiles defined above. Fig.~\ref{fig:bagpipes_age_hist} compares the normalised burst age distributions in each. The youngest sample is weighted towards younger burst ages, while the oldest sample is shifted towards systematically older ages, with the median age (dotted vertical lines) increasing from the youngest to the oldest sample. We repeat this exercise for the burst mass fraction ($f_{\rm burst,50}$) and the quenching timescale ($\tau_{\rm quench,50}$), but neither quantity shows a comparably clear or consistent trend along the $u$ direction in this subsample; any apparent variation is not robust to the choice of binning, likely a consequence of the modest sample size available for this check. Position along $u$ therefore tracks burst age rather than burst strength in this subsample, supporting our interpretation of the sub-samples as an ageing sequence.

\subsubsection{Evolution of CGM absorption within the post-starburst population}
\label{sec:results_psb}

\begin{figure*}
  \centering
  \includegraphics[width=\textwidth]{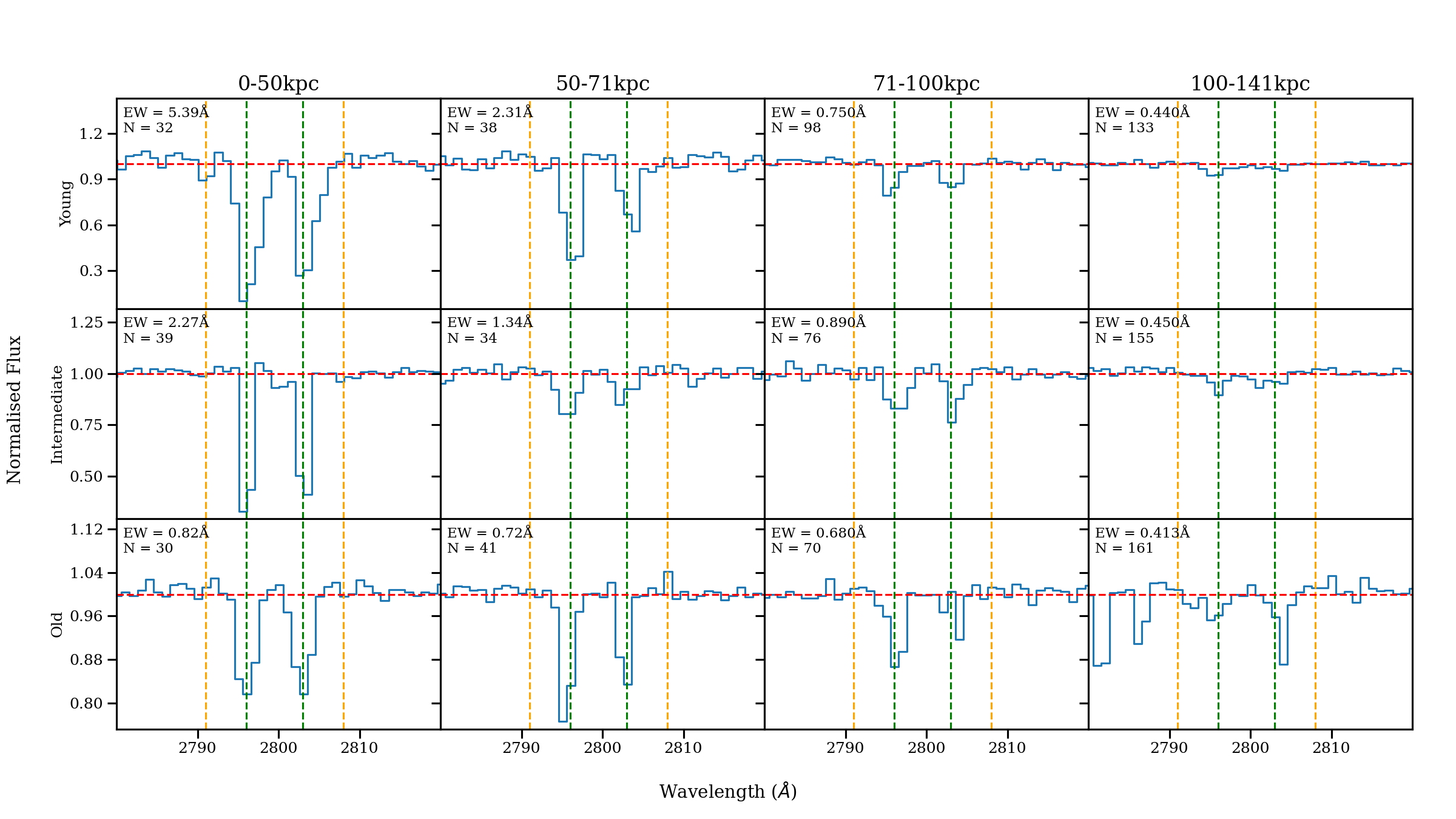}
  \caption{Stacked \mgii\ absorption profiles for the three post-starburst sub-samples, for the innermost four impact parameter bins. Rows correspond to increasing time since the starburst event, from youngest (top) to oldest (bottom). As time since the burst increases, absorption at small radii decreases while absorption at intermediate radii strengthens, consistent with outward redistribution of cool gas. Panels within each row share a common $y$-axis range so that absorption depths can be compared directly across impact parameter - equivalent widths are measured by direct integration over both doublet lines.}
  \label{fig:psbstacks}
\end{figure*}

\begin{figure*}
  \centering
  \includegraphics[width=\textwidth]{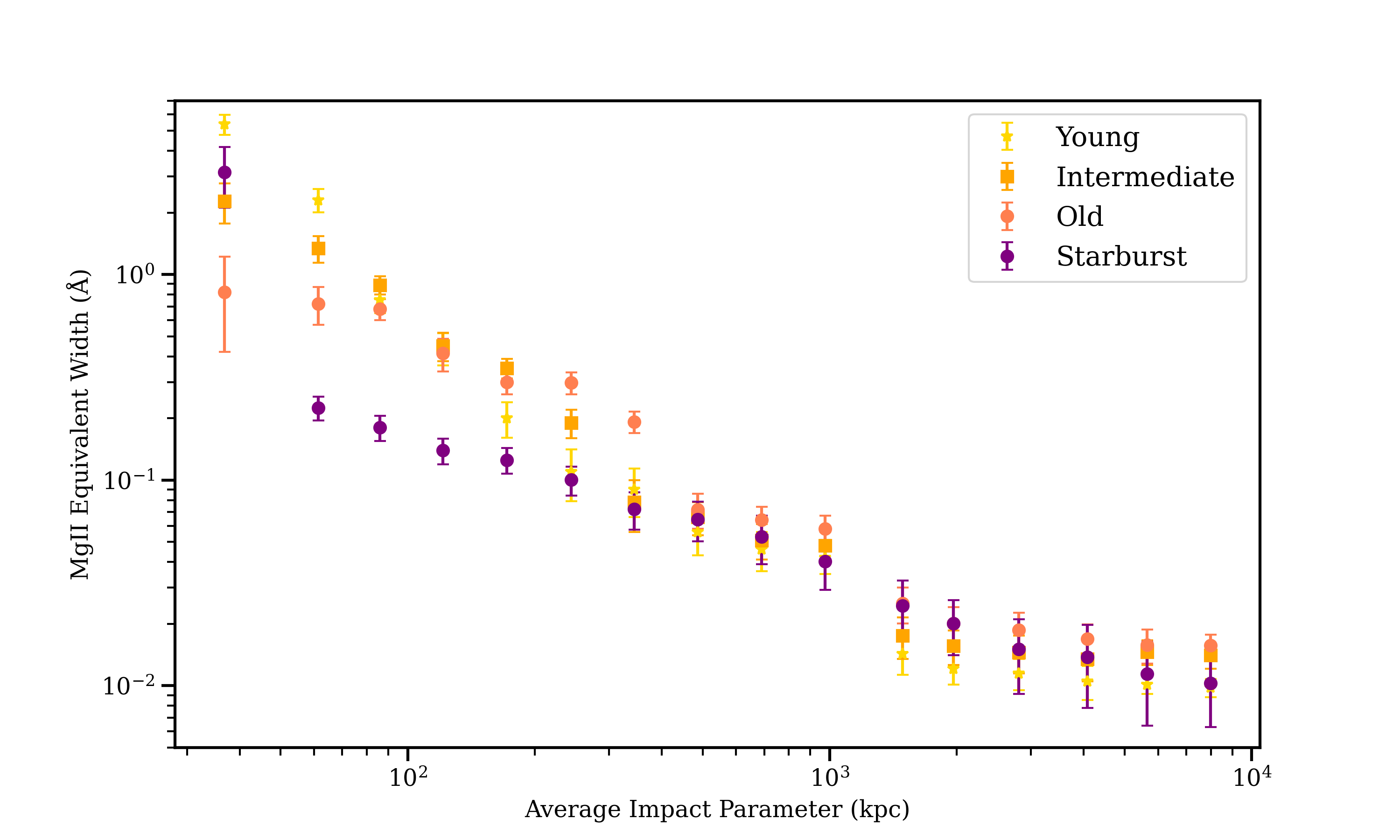}
  \caption{The \mgii\ equivalent width as a function of average impact parameter for the three post-starburst sub-samples, ordered by increasing time since the starburst event, from youngest to oldest. Linear fits are not included in this figure in order to show the step decreases between sub-samples. The youngest post-starburst systems show the steepest inner profiles and the highest amplitude at small impact parameters; the profiles flatten and the inner--outer contrast diminishes as the population ages.}
  \label{fig:eqw_psb_quartiles}
\end{figure*}

The key result of this paper is shown in Figs~\ref{fig:psbstacks} and~\ref{fig:eqw_psb_quartiles}. Sub-dividing the post-starburst population into three equal-sized samples, shown in Table \ref{tab:psb_pairs} ordered by inferred time since the starburst, from the youngest to the oldest, reveals clear and systematic evolution in the \mgii\ absorption profile. The galaxy stacks confirming the classification split are shown in Appendix~\ref{app:stacks}. Note that from this point the impact parameter bins are halved: the larger DESI sample size permits a more precise probe of the relationship between impact parameter and equivalent width, so the eight original bins become sixteen.

The youngest post-starburst sub-sample shows the highest equivalent width at small projected separations ($\lesssim 282$~kpc) and the steepest radial decline of all sub-samples. This is consistent with the immediate aftermath of the starburst phase, when feedback-driven outflows have recently expelled large quantities of cool gas, concentrating much of the absorbing material close to the galaxy. As the inferred post-burst age increases, the absorption at small radii decreases while absorption at intermediate radii ($\sim 300$--$1000$~kpc) progressively strengthens. This behaviour is consistent with an outward redistribution of cool circumgalactic gas: material expelled during or immediately after the burst propagates outward with time, shifting the absorption profile to larger projected separations.

\begin{table}[]
    \centering
    \begin{tabular}{c|c|c|c}
          Impact Parameter (kpc) & \multicolumn{3}{c}{No. of Post-Starburst Pairs} \\
          & Youngest & Intermediate & Oldest \\
         \hline
       0 - 50       & 32     & 39     & 30     \\
       50 - 71      & 38     & 34     & 41     \\
       71 - 100     & 98     & 76     & 70     \\
       100 - 141    & 133    & 155    & 161    \\
       141 - 200    & 335    & 320    & 292    \\
       200 - 282    & 633    & 646    & 677    \\
       282 - 400    & 1476   & 1544   & 1467   \\
       400 - 566    & 3162   & 3095   & 3171   \\
       566 - 800    & 6708   & 6693   & 6697   \\
       800 - 1131   & 13842  & 13860  & 13854  \\
       1131 - 1600  & 28079  & 27976  & 27868  \\
       1600 - 2262  & 55638  & 55740  & 55854  \\
       2262 - 3200  & 110628 & 111057 & 111206 \\
       3200 - 4525  & 221945 & 221064 & 221180 \\
       4525 - 6400  & 224219 & 222987 & 223648 \\
       6400 - 9051  & 228654 & 223485 & 222692 \\
    \end{tabular}
    \caption{The number of quasar–galaxy pairs contributing to each post-starburst sub-sample, as a function of projected impact parameter. The post-starburst population is divided into three equal-sized sub-samples ordered by inferred time since the starburst (youngest, intermediate, oldest), using the sixteen halved impact-parameter bins introduced previously. The sub-samples contain approximately equal numbers of galaxies, and the pair counts are correspondingly similar between them in each bin.}
    \label{tab:psb_pairs}
\end{table}

\begin{table*}
\centering
\caption{Mean \mgii\ equivalent width (\AA) as a function of impact parameter for post-starburst galaxies divided into three sub-samples ordered by inferred time since the starburst event, and for starburst galaxies. Errors are from a bootstrap analysis.}
\label{tab:psb_quartile_ew}
\begin{tabular}{c|c|c|c|c}
\hline
Impact Parameter & \multicolumn{4}{c}{Equivalent width (\AA)} \\
(kpc) & Youngest & Intermediate & Oldest & Starburst \\
\hline
0 -- 50      & $5.39 \pm 0.60$   & $2.27 \pm 0.50$   & $0.82 \pm 0.40$   & $3.14 \pm 1.03$   \\
50 -- 71     & $2.31 \pm 0.30$   & $1.34 \pm 0.20$   & $0.72 \pm 0.15$   & $0.225 \pm 0.030$ \\
71 -- 100    & $0.750 \pm 0.100$ & $0.890 \pm 0.090$ & $0.680 \pm 0.080$ & $0.180 \pm 0.025$ \\
100 -- 141   & $0.440 \pm 0.078$ & $0.450 \pm 0.071$ & $0.413 \pm 0.075$ & $0.140 \pm 0.020$ \\
141 -- 200   & $0.200 \pm 0.039$ & $0.350 \pm 0.038$ & $0.300 \pm 0.039$ & $0.125 \pm 0.018$ \\
200 -- 282   & $0.110 \pm 0.031$ & $0.190 \pm 0.030$ & $0.298 \pm 0.036$ & $0.100 \pm 0.016$ \\
282 -- 400   & $0.090 \pm 0.024$ & $0.078 \pm 0.022$ & $0.192 \pm 0.023$ & $0.072 \pm 0.015$ \\
400 -- 566   & $0.056 \pm 0.013$ & $0.066 \pm 0.012$ & $0.072 \pm 0.014$ & $0.065 \pm 0.014$ \\
566 -- 800   & $0.046 \pm 0.010$ & $0.051 \pm 0.010$ & $0.064 \pm 0.010$ & $0.053 \pm 0.014$ \\
800 -- 1131  & $0.042 \pm 0.007$ & $0.048 \pm 0.008$ & $0.058 \pm 0.009$ & $0.0401 \pm 0.011$ \\
\hline
\end{tabular}
\end{table*}

Adopting a stellar-to-halo mass relation
\citep[e.g.][]{Behroozi2019, Moster2018}, the stellar masses of our LRG sample imply approximate virial radii of $\sim700$--$1000$~kpc, although the halo masses of such massive galaxies are not well constrained and we do not attempt a precise estimate here. Projected separations beyond $\sim1$~Mpc therefore lie outside the virial radius of a typical halo, where the measured \mgii\ absorption is expected to be dominated by the two-halo term, that is, by clustered neighbouring galaxies and the larger-scale environment, rather than by gas bound to the target galaxy itself. We therefore restrict our physical interpretation of the circumgalactic medium to separations $\lesssim 1$~Mpc, and regard the signal at larger separations as tracing the galaxy--\mgii\ cross-correlation and the surrounding environment. Within this inner region, the absorption around the oldest post-starburst sub-sample remains systematically elevated relative to star-forming galaxies, implying that residual differences in the inner gas distribution persist throughout the post-starburst phase probed by our sample. At larger separations ($\gtrsim 1$~Mpc) the profiles of the oldest sub-sample and the star-forming population converge, consistent with the two populations inhabiting comparable large-scale environments.

\section{Discussion}\label{sec:discussion}

\subsection{\mgii\ absorption around post-starburst galaxies}\label{sec:disc:psb}

Post-starburst galaxies exhibit the strongest cool-gas absorption of any spectral class across almost the full range of impact parameters probed (Fig.~\ref{fig:eqw_b_class}), reaching $W_{\rm MgII} = 1.58 \pm 0.32$~\AA\ within the innermost bin ($<71$~kpc). This exceeds the absorption around star-forming ($0.69 \pm 0.15$~\AA) and quiescent ($0.22 \pm 0.05$~\AA) galaxies of comparable stellar mass at the same projected separation. The enhancement confirms and extends earlier evidence, from ultraviolet absorption-line studies of individual sightlines, that post-starburst galaxies retain large reservoirs of cool circumgalactic gas after the cessation of star formation \citep{Borthakur2013,Heckman2017,Harvey2025}, and establishes this as a statistically robust property of the population as a whole.

The evolution of the absorption profile within the post-starburst phase (Figs.~\ref{fig:psbstacks} and \ref{fig:eqw_psb_quartiles}) provides a more detailed physical picture. Our central result is that the evolutionary stage of the Post-Starburst Galaxy correlates with the raised level of cool gas in their CGM. The youngest sub-sample shows both the highest values at small separations ($\lesssim 282$~kpc) and the steepest radial decline, while progressively older sub-samples show reduced absorption at small radii and enhanced absorption at intermediate radii ($\sim 300$--$1000$~kpc). Taken together, this can be interpreted as the outward redistribution of cool gas expelled during or shortly after the starburst: material launched into the halo by feedback-driven winds propagates to larger distances as the galaxy ages, shifting the absorption profile outward while the inner reservoir is gradually depleted. We have verified that these differences between spectral classes are not driven by the differing stellar-mass and redshift distributions of the samples by matching and mass-split testing, presented in Appendix \ref{sec:appendixB}, which confirms that the absorption strength correlates with galaxy classification. 

The independent \textsc{Bagpipes} analysis is important in this context, as it confirms that our MFICA-based ordering corresponds to a sequence in time since the burst rather than to some other spectral property: the median burst age increases along the sub-sample ordering. The profile evolution described above can therefore be interpreted as an ageing sequence. A simple timescale argument supports the outflow interpretation. Cool gas travelling at a characteristic velocity $v$ for a time $t$ reaches a radius of $r \sim v\,t$. Down-the-barrel studies of post-starburst outflows report velocities from a few hundred up to $\sim 1000$~km\,s$^{-1}$ \citep{tremontiOriginPoststarburstGalaxies2007}. For an outflow velocity of $\sim 350$~km\,s$^{-1}$ acting over the
$\sim\!1$~Gyr since the burst, the expelled gas travels $\sim 350$~kpc, comparable to the radii at which we observe the intermediate-scale enhancement in the older sub-samples. The observed migration of the absorption signal is thus consistent with plausible wind velocities and quenching timescales for this population.

We caution that the \mgii\ equivalent width is not a direct measure of gas mass, since the doublet is typically saturated; the profiles above are more properly read as tracing the covering fraction and radial distribution of cool gas rather than its total quantity. Nonetheless, the systematic radial evolution is difficult to reconcile with a static reservoir. Finally, we note that even the oldest post-starburst sub-sample retains absorption in excess of the quiescent population at small radii ($\lesssim 1$~Mpc), implying that a residual inner cool-gas reservoir persists throughout the post-starburst phase probed here, possibly sustained by recycling or by the slow settling of previously expelled material.

\subsection{Comparison with simulations}
\label{sec:sims}
 
The trends reported here --- the enhanced cool-gas absorption around post-starburst galaxies, its outward migration with time since the burst, and the persistence of a residual inner reservoir in the oldest systems - provide observational constraints that can be compared, at least qualitatively, with the picture of feedback-regulated circumgalactic gas that emerges from modern galaxy-formation simulations. Both cosmological and zoom-in simulations predict that episodes of intense star formation drive metal-enriched outflows that transport cool gas from the interstellar medium into the surrounding halo, producing a multiphase circumgalactic medium whose properties depend on the recent star-formation and feedback history of the galaxy \citep{Muratov2015, Nelson2020, FaucherGiguere2023}.
 
The observed evolution of the absorption profile within the post-starburst phase is broadly consistent with this scenario. In the FIRE simulations, stellar feedback drives bursty, high-velocity outflows whose launching is closely tied to peaks in the star-formation rate \citep{Muratov2015}, so that a rapidly truncated burst is naturally expected to be followed by a wave of cool gas propagating outward through the halo. For the characteristic wind velocities of a few hundred $\mathrm{km\,s^{-1}}$ inferred for post-starburst outflows, such material reaches intermediate halo radii of order several hundred kiloparsecs on $\sim$Gyr timescales, comparable to the scale at which we observe the absorption enhancement to shift in the older sub-samples. Simulations that resolve the small-scale structure of the cool CGM similarly find that cool clouds can survive and be distributed over a substantial fraction of the halo \citep{Nelson2020}, providing a plausible physical basis for the extended, radially evolving cool-gas distribution traced by our stacks. We caution, however, that this comparison is qualitative: a quantitative test would require forward-modelling Mg\,{\sc ii} absorption in simulated haloes selected to match our post-starburst sample, which few existing simulations provide.
 
The persistence of an inner cool-gas excess in even the oldest post-starburst sub-sample is also of interest in this context. Simulations that follow the baryon cycle in detail predict that a significant fraction of expelled gas does not escape the halo but is re-accreted through galactic-fountain and recycling flows on timescales of up to a few Gyr \citep{Muratov2015, FaucherGiguere2023}, and that cool gas can (re-)form in situ through thermal instability in the hot halo \citep{Nelson2020}. Either mechanism could sustain the residual inner reservoir we infer, and would be consistent with the absorption at small radii remaining elevated above the quiescent level long after the burst. Distinguishing between these possibilities - and establishing whether the observed migration  reflects a single outflow episode or a more continuous cycle of ejection and return - will require simulations that both track cool-gas tracers explicitly and isolate the post-starburst population, and we regard this as a promising direction for future work.

\section{Summary and Conclusions}
\label{sec:conclusions}

We have measured the cool circumgalactic medium around massive post-starburst galaxies, traced by \mgii\ $\lambda\lambda2796,2803$ absorption in background quasar spectra from DESI DR2. Cross-matching 1,975,891 quasar sightlines with 2,239,234 luminous red galaxies at $0.4 < z < 0.76$ yields approximately 143 million galaxy--quasar pairs, 50--60 times more per impact-parameter bin than the SDSS-based analysis of \citet{Harvey2025}. Foreground galaxies are classified using MFICA, and the post-starburst population is further divided into three sub-samples ordered by time since the starburst. Our main findings are as follows.

\begin{enumerate}

\item The mean \mgii\ equivalent width declines monotonically with impact parameter for every spectral class. Absorption is detected from within 71~kpc out to the largest separations probed ($\sim 9$~Mpc), and a redshift-randomisation test confirms that the signal is physically associated with the foreground galaxies rather than arising from noise or continuum systematics.

\item Post-starburst galaxies show the strongest \mgii\ absorption of any spectral class across almost the full range of impact parameters, reaching $W_{\rm MgII} = 1.58 \pm 0.32$~\AA\ within 71~kpc. This is above star-forming ($0.69 \pm 0.15$~\AA) and quiescent ($0.22 \pm 0.05$~\AA) galaxies of comparable stellar mass at the same projected separation. Starburst galaxies show comparably strong absorption at the smallest separations but a steeper radial decline. Beyond $\sim 1$~Mpc the profiles of all five classes converge.

\item Within the post-starburst phase the absorption profile evolves systematically. The youngest sub-sample shows the highest and most centrally concentrated absorption and the steepest radial decline, while progressively older sub-samples show weaker absorption at small radii and enhanced absorption at intermediate radii ($\sim 300$--$1000$~kpc).

\item Independent \textsc{Bagpipes} star formation history fitting of a high signal-to-noise post-starburst subsample confirms that the MFICA-based ordering corresponds to a sequence in time since the burst: the median burst age increases from the youngest to the oldest sub-sample. The sub-division can therefore be read as an ageing sequence rather than as a trend in some other spectral property.

\item We interpret the evolution of the profile as the outward redistribution of cool gas expelled during or shortly after the starburst. For a characteristic outflow velocity of a few hundred $\mathrm{km\,s^{-1}}$ acting over the $\sim 1$~Gyr since the burst, expelled material travels a few hundred kiloparsecs, comparable to the radii at which the intermediate-scale enhancement is observed in the older sub-samples.

\item Even the oldest post-starburst sub-sample retains absorption in excess of the quiescent population at small radii, implying that a residual inner cool-gas reservoir persists throughout the post-starburst phase probed here. This may be sustained by recycling, by the slow settling of previously expelled material, or by in-situ formation of cool gas through thermal instability in the hot halo.

\item These differences are not driven by the differing stellar-mass and redshift distributions of the samples. Matching and mass-split tests (Appendix~\ref{sec:appendixB}) show that absorption strength correlates with recent star formation history at fixed impact parameter.

\end{enumerate}

Two caveats bound this interpretation. The \mgii\ equivalent width is not a direct measure of gas mass, since the doublet is typically saturated, so the profiles are better read as tracing the covering fraction and radial distribution of cool gas than its total quantity. In addition, projected separations beyond $\sim 1$~Mpc lie outside the virial radius of a typical halo in our sample, where the measured absorption is expected to be dominated by the two-halo term, and we have restricted our physical interpretation accordingly.

The picture that emerges is one in which the cool circumgalactic gas reservoir of a post-starburst galaxy is not static but continues to evolve for at least a Gyr after star formation has ceased, with material migrating outward through the halo while an inner reservoir persists. Distinguishing whether this reflects a single outflow episode or a more continuous cycle of ejection and return will require simulations that track cool-gas tracers explicitly and isolate the post-starburst population. Observationally, extending this analysis to other ions, and to the still larger samples expected from later DESI data releases, would allow the phase structure and metal content of the evolving reservoir to be constrained directly.

\section*{Acknowledgements}

We would like to acknowledge the contribution of Paul C Hewett, whose work underpins the MFICA classification and analysis in this work. 

This material is based upon work supported by the U.S. Department of Energy (DOE), Office of Science, Office of High-Energy Physics, under Contract No. DE–AC02–05CH11231, and by the National Energy Research Scientific Computing Center, a DOE Office of Science User Facility under the same contract. Additional support for DESI was provided by the U.S. National Science Foundation (NSF), Division of Astronomical Sciences under Contract No. AST-0950945 to the NSF’s National Optical-Infrared Astronomy Research Laboratory; the Science and Technology Facilities Council of the United Kingdom; the Gordon and Betty Moore Foundation; the Heising-Simons Foundation; the French Alternative Energies and Atomic Energy Commission (CEA); the Secretariat of Science, Humanities, Technology and Innovation (SECIHTI) of Mexico; the Ministry of Science, Innovation and Universities of Spain (MICIU/AEI/10.13039/501100011033), and by the DESI Member Institutions: \url{https://www.desi.lbl.gov/collaborating-institutions}. Any opinions, findings, and conclusions or recommendations expressed in this material are those of the author(s) and do not necessarily reflect the views of the U. S. National Science Foundation, the U. S. Department of Energy, or any of the listed funding agencies.

RT is supported by the Science and Technologies Facilities Council (STFC, grant number UKRI4102). VW acknowledges Science and Technologies Facilities Council (STFC) grant ST/Y00275X/1 and Leverhulme Fellowship RF-2024-589/4.

The authors are honored to be permitted to conduct scientific research on I'oligam Du'ag (Kitt Peak), a mountain with particular significance to the Tohono O’odham Nation.

\section{Data Availability}

This work is based on data from Data Release 2 (DR2) of the Dark Energy Spectroscopic Instrument. At the time of writing, the DR2 spectra and redshifts have not yet been publicly released; they will be made available in a future public DESI data release, in accordance with the DESI data management plan. DESI data are distributed under the Creative Commons Attribution 4.0 International licence. The Mean-Field Independent Component Analysis classifications will be published as a Value Added Catalogue as part of DESI's DR2. The data corresponding to the figures in this paper will be available upon acceptance. The data corresponding to the figures in this paper will be available upon acceptance.

The value-added products generated for this analysis - including the galaxy-quasar pair catalogue, the \textsc{Bagpipes} posterior estimates for the post-starburst subsample, the stacked Mg\,{\sc ii} spectra, and the derived equivalent-width measurements - are available from the corresponding author on reasonable request.

\bibliography{DESIPaper}



\appendix

\section{Composite spectra of the galaxy classes and post-starburst sub-samples}
\label{app:stacks}
 
To illustrate the spectral basis of the MFICA classification introduced in Section~\ref{sec:data_specclass}, we show mean-stacked rest-frame spectra of the foreground DESI LRG galaxies, co-added within each spectral class and within each post-starburst sub-sample. The individual galaxy spectra are shifted to the rest frame, resampled onto a common wavelength grid spanning $3300$--$5200$~\AA\ (the range used in the MFICA decomposition), and normalised over a common continuum window (4150--4250~\AA) before stacking.
 
\begin{figure*}
    \centering
    \includegraphics[width=\textwidth]{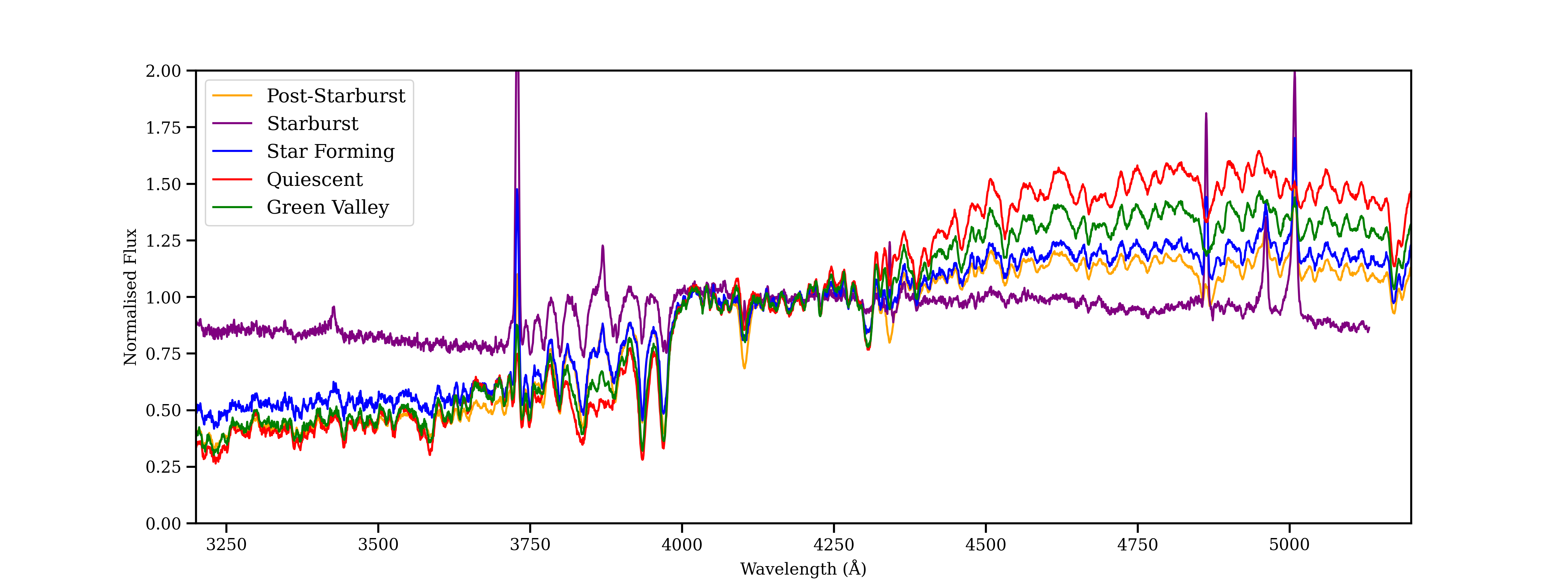}
    \caption{Mean-stacked rest-frame spectra of the foreground DESI LRG galaxies in each of the five MFICA spectral classes: post-starburst (orange), starburst (purple), star-forming (blue), quiescent (red) and green valley (green). The starburst composite shows a blue continuum, a weak $4000$~\AA\ break and strong nebular emission, while the quiescent composite shows a red continuum and a strong break; the post-starburst composite is distinguished by deep higher-order Balmer absorption. These are the composite spectra underlying the spectral classification used throughout.}
    \label{fig:class_stacks}
\end{figure*}

Figure~\ref{fig:class_stacks} shows the composite spectra of the five principal spectral classes. The sequence from starburst to quiescent traces a systematic change in stellar-population age. The starburst composite has the bluest continuum and the weakest $4000$~\AA\ break, together with strong nebular emission - most visibly [O\,{\sc ii}]~$\lambda3727$, H$\beta$ and [O\,{\sc iii}]~$\lambda\lambda4959,5007$ - as expected for galaxies dominated by light from young, massive stars during an ongoing burst. The star-forming and green-valley composites show progressively redder continua, stronger $4000$~\AA\ breaks and weaker emission, while the quiescent composite has the reddest continuum, the strongest $4000$~\AA\ break and prominent Ca\,{\sc ii}~H\&K absorption with no detectable emission, consistent with an old, passively evolving stellar population. The post-starburst composite is intermediate in continuum shape but is set apart by conspicuously deep higher-order Balmer absorption (H$\delta$, H$\gamma$ and the Balmer series near $3750$--$3950$~\AA), the characteristic signature of the A- and F-type stars that dominate the light shortly after a rapidly truncated burst. It is precisely this feature that the intermediate-age (``AF'') MFICA component is sensitive to, and that underlies the separation of post-starburst galaxies from the other classes.
 
\begin{figure*}
    \centering
    \includegraphics[width=\textwidth]{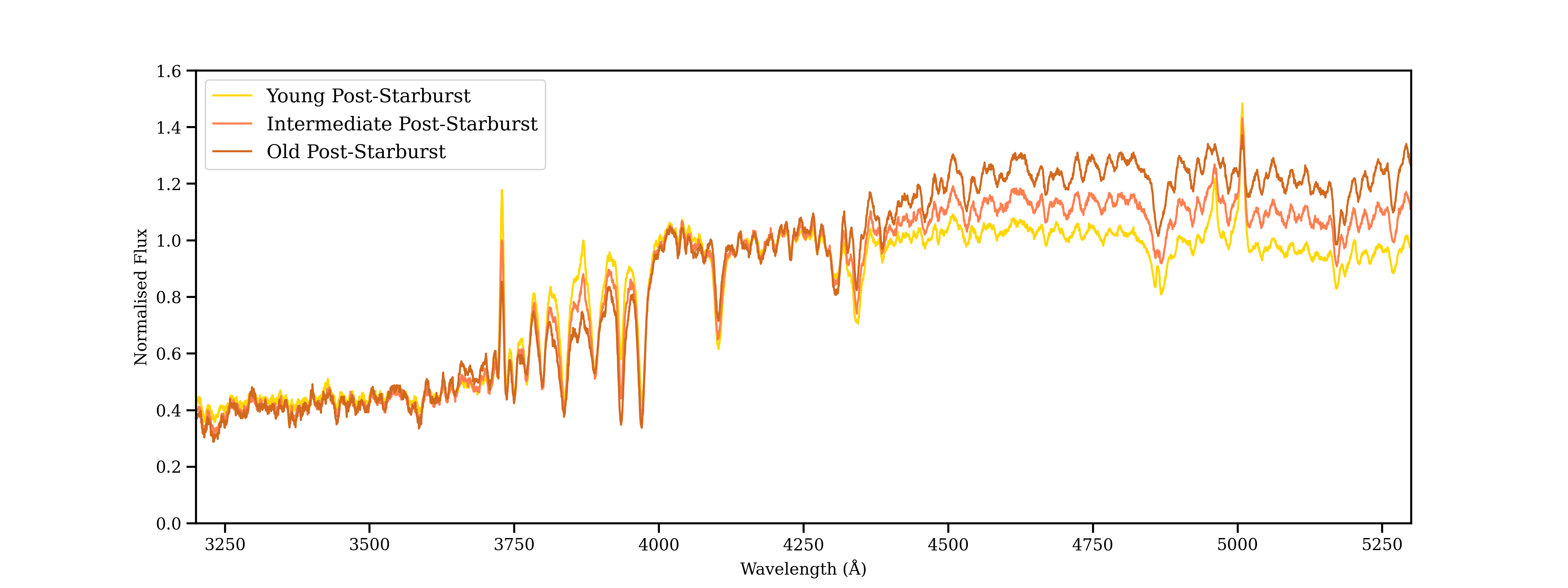}
    \caption{Mean-stacked rest-frame spectra of the three post-starburst sub-samples, ordered by inferred time since the starburst: young (yellow), intermediate (salmon) and old (brown). From young to old the continuum reddens, the $4000$~\AA\ break strengthens and the Balmer absorption weakens, consistent with an ageing stellar population. These are the composite spectra underlying the post-starburst sub-division used throughout.}
    \label{fig:psb_stacks}
\end{figure*}

Figure~\ref{fig:psb_stacks} shows the composite spectra of the three post-starburst sub-samples, ordered by their inferred time since the starburst (``young'', ``intermediate'' and ``old''; Section~\ref{sec:data_specclass}). A clear and monotonic spectral progression is apparent: from the youngest to the oldest sub-sample the continuum reddens, the $4000$~\AA\ break strengthens, and the depth of the Balmer absorption decreases as the A/F stellar population fades. The youngest sub-sample retains the bluest continuum and the strongest Balmer absorption, as expected soon after quenching, whereas the oldest sub-sample is markedly more quiescent-like. This trend is consistent with an ageing sequence and provides a direct, model-independent visual counterpart to the increase in BAGPIPES burst age along the sub-sample ordering, supporting the interpretation of the MFICA sub-division as a proxy for time since the starburst.
 
\section{The impact of redshift and stellar mass on equivalent width versus impact parameter}
\label{sec:appendixB}

The five spectral classes differ in their redshift and stellar-mass distributions; in particular, the starburst sample extends to lower stellar masses than the other classes. Because \ion{Mg}{2} equivalent width may depend on both quantities, we test here whether the differences between spectral classes reported in Section~\ref{sec:results_class} are driven by recent star-formation history rather than by these underlying population differences. This repeats, for the substantially larger DESI DR2 sample, the analysis presented in Appendix~B of \citet{Harvey2025}.

Figure~\ref{fig:appendixB} shows the equivalent width versus impact parameter relation for each spectral class, with each class split at its median redshift (top row) and at its median stellar mass (bottom row). The left-hand panels compare the star-forming, quiescent and green valley samples, and the right-hand panels compare the star-forming, post-starburst and starburst samples.

\textit{Redshift.} Over the redshift range studied here ($0.4 < z < 0.76$), the high- and low-redshift relations lie close to the full-sample relation for every class, and in most cases within its uncertainty. The largest offsets, in the inner normalisation of the green valley sample and in the star-forming sample beyond $\sim$5~Mpc, are small compared with the separation between classes. There is no strong trend between \ion{Mg}{2} equivalent width and redshift for these very massive galaxies, and the gradients of the relations are not significantly altered. The ordering of the classes is preserved in both redshift subsamples. Within $\sim$300~kpc, post-starburst galaxies show the strongest absorption, followed by star-forming, green valley and quiescent galaxies. Starburst galaxies show a flatter profile, with weaker absorption than star-forming galaxies at small impact parameters and stronger absorption beyond $\sim$500~kpc.

We additionally recalculate the stacks by (a) weighting the input spectra of the comparison subsamples to match the redshift distribution of the post-starburst galaxies, and (b) sub-sampling the comparison samples to have redshift distributions matching that of the post-starburst galaxies. In both cases there is no qualitative change in our results. The sub-sampled version also provides an independent check on the bootstrap errors in bins with few objects, since each stack can be drawn from a larger pool of spectra than is used to construct it. As expected, the errors increase slightly, by at most $\sim$10~per~cent, without affecting the overall result.

\textit{Stellar mass.} To test whether our results might change qualitatively if we normalised by a stellar-mass-scaled quantity such as the halo virial radius, we split each class at its median stellar mass (Figure~\ref{fig:appendixB}, bottom row). The high- and low-mass relations are almost indistinguishable from the full-sample relation for all five classes, agreeing even more closely than the redshift subsamples. Mass therefore does not determine absorption strength within this high-mass galaxy sample. In particular, the distinct profile of the starburst class is unchanged in both stellar-mass subsamples. This confirms that it is not an artefact of the lower characteristic stellar mass of the starburst population.

Regardless of method, absorption strength at fixed impact parameter correlates strongly with the recent star-formation history of galaxies, and not with their redshift or stellar mass.

\begin{figure*}
    \centering
    \includegraphics[width=\textwidth]{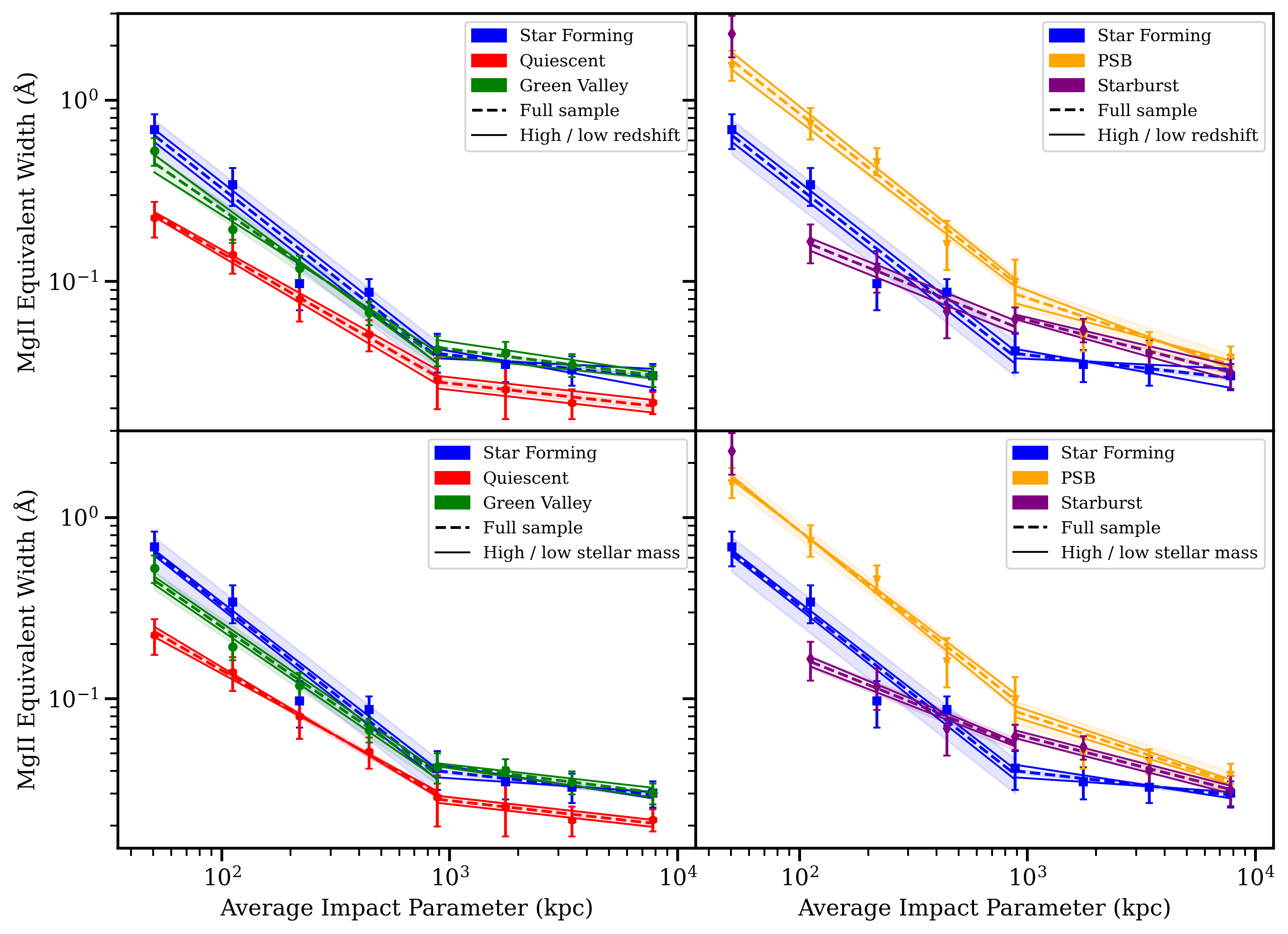}
    \caption{\ion{Mg}{2} equivalent width as a function of average impact parameter, with each spectral class split at its median redshift (top row) and median stellar mass (bottom row). Left: star-forming (blue), quiescent (red) and green valley (green) galaxies. Right: star-forming (blue), post-starburst (PSB; orange) and starburst (purple) galaxies. Points with error bars show the stacked measurements for the full sample of each class. Dashed lines show the best-fitting relation for the full sample, with the shaded region indicating its uncertainty, and solid lines show the best-fitting relations for the high and low redshift (or stellar mass) subsamples. In all cases the subsample relations closely follow the full-sample relation, and the differences between spectral classes are preserved.}
    \label{fig:appendixB}
\end{figure*}

\section{Example \textsc{Bagpipes} fit to a post-starburst galaxy}
\label{sec:appendixC}
 
To illustrate the star-formation-history (SFH) fitting underlying the independent age validation, we show here the full \textsc{Bagpipes} \citep{Carnall2018} fit to a single representative post-starburst galaxy from the high-signal-to-noise subsample ($z = 0.51$). The fit uses the two-component ``\texttt{psb two-step}'' SFH model of \citet{Leung2024}, in which an old, exponentially declining stellar population is combined with a more recent burst, together with a dust attenuation law and a fitted spectral noise and velocity-dispersion model.
 
\begin{figure*}
    \centering
    \includegraphics[width=\textwidth]{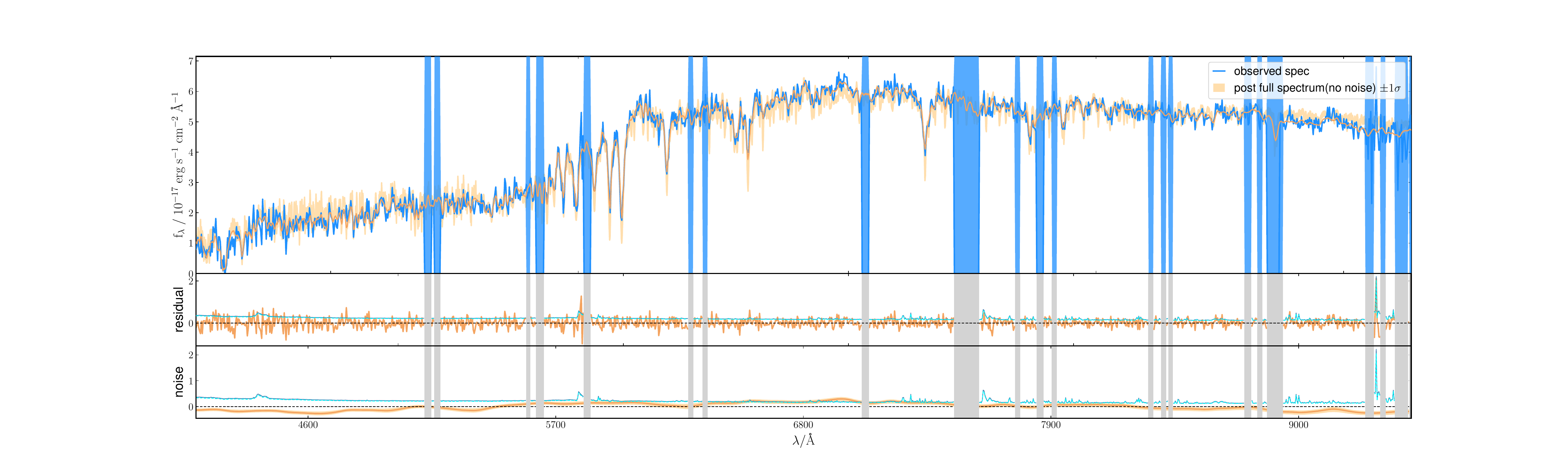}
    \caption{\textsc{Bagpipes} fit to the spectrum of a representative post-starburst galaxy at $z = 0.51$. \emph{Top:} the observed DESI spectrum (blue) with the posterior median model spectrum and its $1\sigma$ credible interval overlaid. \emph{Middle:} fit residuals, showing no significant systematic structure. \emph{Bottom:} the fitted noise model. The model reproduces both the continuum shape and the deep higher-order Balmer absorption characteristic of the post-starburst class.}
    \label{fig:appC_fit}
\end{figure*}

Figure~\ref{fig:appC_fit} shows the observed DESI spectrum together with the posterior median model spectrum and its $1\sigma$ credible interval. The model reproduces both the overall continuum shape and the strong higher-order Balmer absorption that characterises the post-starburst population, with the residual panel showing no significant systematic structure across the fitted wavelength range. The lower panel shows the fitted noise model.
 
\begin{figure*}
    \centering
    \includegraphics[width=\textwidth, trim=0 0 0 0, clip]{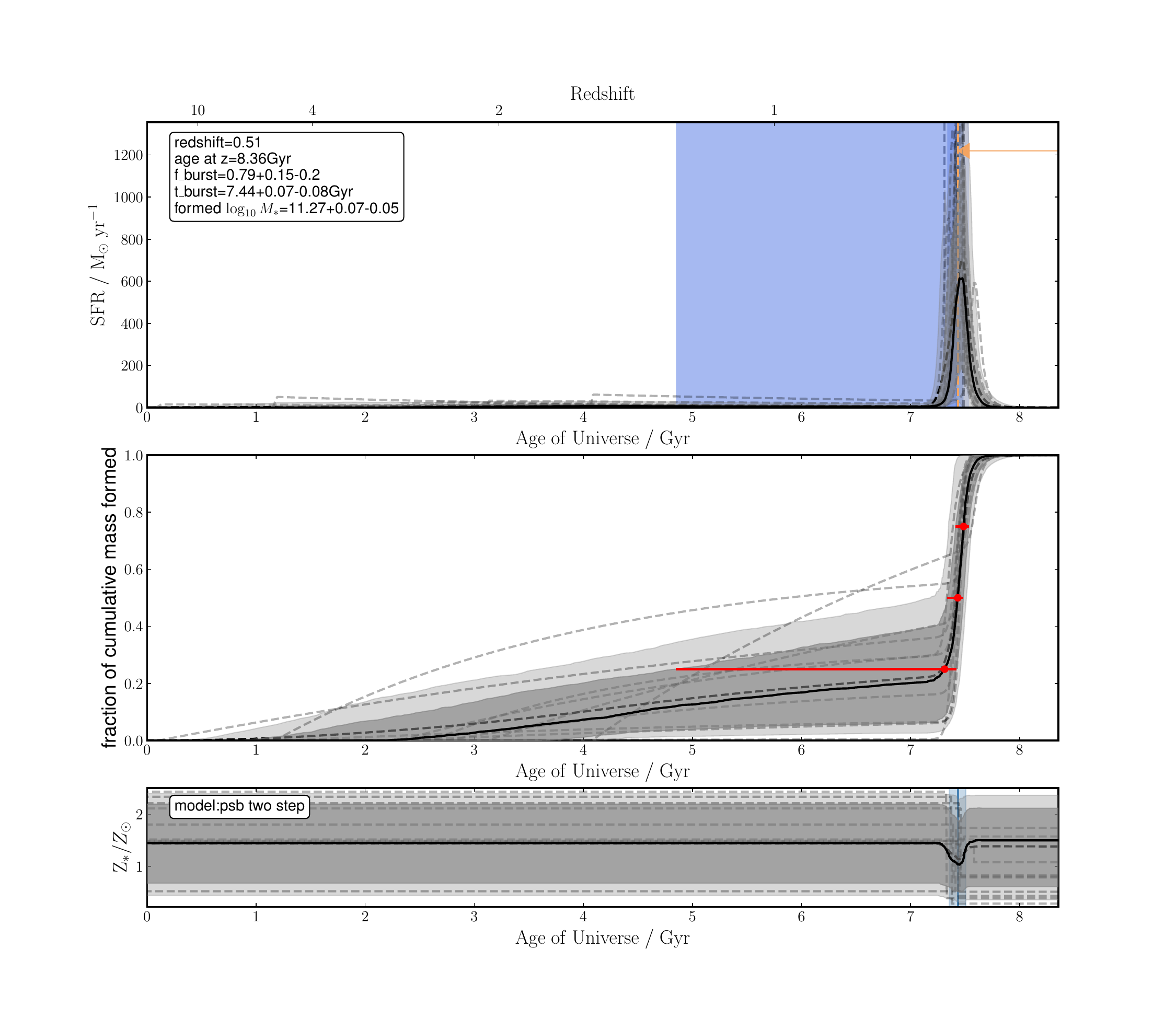}
    \caption{Inferred star-formation history for the same galaxy, using the two-component \texttt{psb two-step} model of \citet{Leung2024}. \emph{Top:} star-formation rate versus age of the Universe (upper axis: redshift), with the posterior median (black) and credible intervals (grey); the fit recovers a sharp, recent, and rapidly truncated burst. \emph{Middle:} cumulative fraction of stellar mass formed, with the $16$th, $50$th and $84$th percentiles marked (red). \emph{Bottom:} stellar metallicity. Summary parameters for this galaxy are annotated in the top panel.}
    \label{fig:appC_sfh}
\end{figure*}

Figure~\ref{fig:appC_sfh} shows the inferred SFH. The top panel gives the star-formation rate as a function of the age of the Universe, the middle panel the cumulative fraction of stellar mass formed, and the bottom panel the stellar metallicity. The reconstructed history is characteristic of a post-starburst system: a low level of early star formation followed by a sharp, recent burst that dominates the mass assembly and is then rapidly truncated. For this galaxy the fit recovers a burst that formed a fraction $f_{\mathrm{burst}} = 0.79^{+0.15}_{-0.20}$ of the stellar mass, a total formed stellar mass of $\log_{10}(M_*/M_\odot) = 11.27^{+0.07}_{-0.05}$, and a burst occurring at an age of the Universe of $t_{\mathrm{burst}} = 7.44^{+0.07}_{-0.08}\,\mathrm{Gyr}$ (a lookback burst age of $0.93^{+0.08}_{-0.07}\,\mathrm{Gyr}$ at the redshift of the galaxy). These values place the galaxy firmly within the recently quenched regime for which the model parametrisation is appropriate.
 
The burst age, burst mass fraction and stellar mass are well constrained, whereas the quenching timescale ($\tau_{\rm psb} = 5.26^{+3.21}_{-3.14}$\,Gyr) is poorly determined, as expected from a single moderate-resolution spectrum. We show this single object as a representative example; the burst ages derived from fits of this kind, applied to the full high-signal-to-noise subsample, form the basis of the age ordering presented throughout.
  
\renewcommand\thefigure{\thesection.\arabic{figure}}

\end{document}